\documentclass{article}
\usepackage{xcolor, amsmath, amssymb, amsthm, bm, multirow, url, graphicx, comment, caption, subcaption, booktabs}
\usepackage[authoryear]{natbib}
\usepackage[british]{babel}
\usepackage[useregional]{datetime2}
\DTMlangsetup[en-GB]{showdayofmonth=false}
\usepackage{authblk}
\usepackage[colorlinks,citecolor=blue,urlcolor=blue,hypertexnames=false]{hyperref}

\usepackage[linesnumbered, ruled, noline]{algorithm2e}
\SetKwInput{KwInit}{Initialization}
\SetKwInput{KwInput}{Input}
\SetKwInput{KwOutput}{Output}

\usepackage{setspace}

\theoremstyle{plain}

\newtheorem{dgp}{DGP}

\title{\textbf{Time Series Forecasting with Many Predictors}}
\author{Shuo-Chieh Huang and Ruey S. Tsay \\ 
Booth School of Business, University of Chicago}

\begin{document}
\maketitle

\begin{abstract}
We propose a novel approach for time series forecasting with many predictors, referred to as the GO-sdPCA, in this paper. The approach employs 
a variable selection method known as the group orthogonal greedy algorithm and the high-dimensional Akaike information criterion to mitigate the impact of 
irrelevant predictors. 
Moreover, a novel technique, called peeling, is used to boost the variable selection procedure so that many factor-relevant predictors can be included in 
prediction.
Finally, the supervised dynamic principal component analysis (sdPCA) method is adopted to account for the dynamic information in factor recovery.
In simulation studies, we found that the proposed method adapts well to unknown degrees of sparsity and factor strength, which results in good performance even when the number of relevant predictors is large compared to the sample size.
Applying to economic and environmental studies, the proposed method consistently performs well compared to some commonly used benchmarks in one-step-ahead out-sample forecasts.
\end{abstract}

\section{Introduction}

In the current data-rich environment, many predictors are often available in time series forecasting.
However, conventional methods have encountered serious 
difficulties in exploiting the information contained in such high-dimensional data.
In particular, the curse of dimensionality often leads to unreliable forecasts when the conventional methods are applied naively.
To exploit the information in the high-dimensional predictors, the use of latent factors has emerged among the first successful approaches. See, for instance, \citet{pena1987}, \citet{SW2002,SW2002b}, \citet{lam2011}, and Chapter 6 of \citet{tsay2013multivariate}.
These factor-based methods are also widely used in the econometric analysis of high-dimensional time series \citep[e.g.][]{Bernanke2003, Bernanke2005, Jurado2015, maccraken2016}.

Nevertheless, existing factor-based methods may still be inadequate, and several issues have been constantly observed in practice.
For example, weak factors are prevalent in real-world data and extracting factors using all predictors may not be optimal \citep{Boivin2006}. 
In addition, some factor models may still contain many parameters when the 
factor dimension is not small. On the other hand, 
\citet{Kim2014} reported, from an extensive experiment, that combining shrinkage methods with factor-based approaches can yield more accurate forecasts among many targeted macroeconomic time series. This implies that there might exist 
some irrelevant predictors in an empirical application. 
The recently proposed scaled principal component analysis of \citet{Huang2022} and the supervised dynamic PCA (sdPCA) of \citet{gao2023} partially remedy this issue by constructing factors in a supervised fashion, so that the effects of irrelevant predictors are reduced. Limited experience indicates that when the number of predictors far exceed the sample size or when the predictors are highly correlated, the performance of such supervised approaches can still be compromised.

To efficiently extract predictive factors from high-dimensional data, in this paper, we propose a new method that blends variable selection in factor estimation.
The proposed method uses the group orthogonal greedy algorithm (GOGA, \citealt{chan2017}) to screen variables that are relevant to the prediction problem at hand, and apply the sdPCA of \citet{gao2023} to extract factors from the selected variables.
The orthogonal greedy algorithm, as well as its variant GOGA for grouped predictors, have been employed for variable selection for high-dimensional linear models, especially with dependent data and highly correlated predictors \citep{ing2020, chan2017, ing2011}.
Since both GOGA and sdPCA serve to minimize the effects of noisy irrelevant variables, the combined procedure, which we call GO-sdPCA, can effectively construct highly predictive factors.

It is worth pointing out that the novelty of our method lies in a key step that successfully combines variable selection with factor-based methods.
Indeed, variable selection techniques commonly employed such as the Lasso of \citet{tib1996} or the OGA of \citet{ing2011} tend to select a sparse representation of the data.
These methods will select only a few variables when many variables are driven by common factors and thereby highly correlated.
For factor estimation, this can be undesirable since factor recovery typically benefits from the many variables that are loaded on the shared factors \citep[see, e.g.][]{lam2011, gao2023}.
This issue can be even more pronounced when the factors are strong, in which case the factors can be accurately recovered by employing many predictors. 
To circumvent this predicament, we propose to use a ``peeling'' technique, which repeatedly apply GOGA to the data after previously selected variables are dropped from the set of candidate predictors.
In this way, our method can select more variables in the model and discriminate between relevant and irrelevant predictors.
In Section \ref{Sec4}, we intoduce the proposed method in more details.
To better motivate the proposed method, we briefly review the orthogonal greedy algorithm (OGA) and its variants in Section \ref{Sec2}. 
We also review the high-dimensional information criteria that are often used to balance the model complexity and the model fit along OGA iterations.
The sdPCA is reviewed in Section \ref{Sec3}, where we particularly emphasize the method's design principles.

We use simulation studies and some empirical analyses to examine the 
performance of the proposed GO-sdPCA approach.
In addition to comparing with some factor-based methods, such as the diffusion index approach of \citet{SW2002, SW2002b}, the sdPCA, and the time series factor model of \citet{lam2011}, we also compare GO-sdPCA against the Lasso and the random forests of \citet{Breiman2001}, both of which are widely-used and versatile tools in machine learning. 
We found that GO-sdPCA improves upon most of factor-based approaches and 
fares well with selected competing methods, even when the number of relevant variables is comparable with the sample size. 
The simulation results are discussed in Section \ref{Sec5}.

In Section \ref{Sec6}, we apply the proposed method to two real datasets.
The first dataset  is the well-known FRED-MD macroeconomic data \citep{maccraken2016}.
Time series forecasting has been a vital topic both in econometric research and in policy-making pertaining to this data set.
The other dataset contains the hourly particulate matters (PM$_{2.5}$) measurements in Taiwan.
The PM$_{2.5}$ data play an important role in the environmental studies as well as informing public health policies.
However, since its measurements are taken over time and space, the dataset  is naturally of high dimension and presents serious challenges to forecast. 
Our findings demonstrate that the proposed method consistently offers more accurate forecasts than the competing methods, validating its practical utility.

\section{Orthogonal greedy algorithm} \label{Sec2}

In this section, we briefly review the (group) orthogonal greedy algorithm. 
The OGA is an iterative algorithm that sequentially chooses predictors to form a regression model. 
Theoretically grounded in approximation theory \citep{Temlyakov2011}, the OGA is also easy to implement computationally.
In the statistical learning and high-dimensional linear regression literature, \citet{barron2008} and \citet{ing2011} analyzed its convergence rate as well as variable screening capability. Its generalization to grouped predictors was studied by \citet{chan2017} with application to threshold autoregressive time series models. 

Throughout this paper, we denote by $\mathbf{y} = (y_{1}, \ldots, y_{n})^{\top}$ the data of the response variable we are interested in, where $n$ is the sample size.
We also have data for the $p$ predictors, $\{\mathbf{x}_{(j)}: j = 1,2, \ldots, p\}$, where $\mathbf{x}_{(j)} = (x_{1,j}, \ldots, x_{n,j})^{\top}$.
The OGA is defined as follows.
Starting with $\hat{J}_{0} = \emptyset$ and $\mathbf{u}^{(0)} = \mathbf{y} = (y_{1}, y_{2}, \ldots, y_{n})^{\top}$, the OGA computes, at iteration $k=1,2,\ldots$, 
\begin{align}
    \hat{j}_{k} &= \arg\min_{1 \leq j \leq p} \Vert \mathbf{u}^{(k-1)} - \mathbf{x}_{(j)}(\mathbf{x}_{(j)}^{\top}\mathbf{x}_{(j)})^{-1} \mathbf{x}_{(j)}^{\top} \mathbf{u}^{(k-1)} \Vert^{2}, \label{2-1} \\
\end{align}
and updates
\begin{align*}    
    \hat{J}_{k} &= \hat{J}_{k-1} \cup \{\hat{j}_{k}\}, \notag \\
    \mathbf{u}^{(k)} &= (\mathbf{I} - \mathbf{H}_{(k)})\mathbf{y}. \notag
\end{align*}
where $\mathbf{H}_{(k)}$ is the orthogonal projection matrix associated with the linear space spanned by $\{\mathbf{x}_{(j)}: j\in \hat{J}_{k}\}$.
Clearly, the OGA sequentially selects variable to include in the model; the set $\hat{J}_{k}$ denotes the index set corresponding to the predictors already selected at iteration $k$.
Intuitively, in each iteration OGA selects the variable that best explains the current residuals.
There are numerical schemes to speed up the computation of the residuals such as using sequential orthogonalization. 
We refer to \citet{ing2011} and \citet{chan2017} for details.

For the purpose of this paper, we will make use of a slight generalization of the OGA that deals with grouped predictors. 
\citet{chan2017} studied the group OGA (GOGA) and applied the method to estimate threshold time series models. 
Consider the $j$-th predictor, $\{x_{t,j}\}_{t=1}^{n}$. 
Instead of being a scalar predictor, it contains $d_{j}$ component predictors, $\{(x_{t,j,1}, \ldots, x_{t,j,d_{j}}):t = 1,2,\ldots, n\}$, thereby forming a ``group'' predictor. 
Then, by substituting
\begin{align*}
    \mathbf{x}_{(j)} = \begin{pmatrix}
        x_{1,j,1} & x_{1,j,2} & \ldots & x_{1,j,d_{j}} \\
        x_{2,j,1} & x_{2,j,2} & \ldots & x_{2,j,d_{j}} \\
        \vdots & \vdots & \ddots & \vdots \\
        x_{n,j,1} & x_{n,j,2} & \ldots & x_{n,j,d_{j}}
    \end{pmatrix}
\end{align*}
in \eqref{2-1}, the procedure becomes the GOGA. 

After $K_{n}$ iterations, the GOGA selects the (group) predictors corresponding to $\hat{J}_{K_{n}}=\{\hat{j}_{1}, \hat{j}_{2}, \ldots, \hat{j}_{K_{n}}\}$. 
Bias-variance trade-off manifests when selecting the number of predictors to be included, as a large $K_{n}$ may lead to over-fitting.
To select a desirable model complexity, \citet{ing2020} suggests  using the high-dimensional Akaike information criterion (HDAIC).
The HDAIC of the model at iteration $k$ is defined as
\begin{align} \label{2-2}
    \mathrm{HDAIC}(\hat{J}_{k}) = \left(1 + C \frac{k \log p}{n} \right) \hat{\sigma}_{(k)}^{2},
\end{align}
where $\hat{\sigma}_{(k)}^{2} = n^{-1} \Vert \mathbf{u}^{(k)} \Vert^{2}$, and $C$ is a constant to be tuned. 
Then the model selected by HDAIC is 
\begin{align} \label{2-3}
    \hat{J}_{\hat{k}} = \{\hat{j}_{1}, \ldots, \hat{j}_{\hat{k}}\}, \quad \mbox{where } \hat{k} = \arg\min_{1 \leq k \leq K_{n}} \mathrm{HDAIC}(\hat{J}_{k}).
\end{align}

Theoretically, \citet{ing2020} proved that the resulting model selected by OGA+HDAIC adapts to the underlying sparsity structure. In particular, consider the high-dimensional regression model,
\begin{align*}
    y_{t} = \sum_{j=1}^{p}\beta_{j} x_{t,j} + \epsilon_{t}.
\end{align*}
Then the conditional mean squared prediction error of OGA+HDAIC is of rate
\begin{align*}
    \left\{ \begin{array}{ll}
       \left( \frac{\log p}{n} \right)^{1 - 1/2\gamma},  & \mbox{if } \sum_{j\in J}|\beta_{j}| \leq D_{1} \left( \sum_{j \in J} \beta_{j}^{2} \right)^{(\gamma - 1)/(2\gamma - 1)} \mbox{ for all } J \\
      \frac{\log n \log p}{n},   & \mbox{if } \sum_{j \in J}|\beta_{j}| \leq D_{2} \max_{j \in J} |\beta_{j}| \mbox{ for all } J \\
      \frac{k_{0}\log p}{n}, & \mbox{if } \min_{j:\beta_{j} \neq 0} |\beta_{j}| \geq D_{3},
    \end{array} \right.
\end{align*}
for some $\gamma \geq 1$ and positive constants $D_{1}, D_{2}, D_{3}$. See Theorem 3.1 of \citet{ing2020} for details. 
Note that these rates are minimax optimal, and, are automatically achieved by OGA+HDAIC without knowledge about the sparsity pattern. 
In this paper, we will leverage this property to select good predictors in constructing forecasts. 
In this way, the variables used in our method is carefully selected in a supervised fashion, which is more effective than employing all predictors in the high-dimensional data. 

\section{Supervised dynamic PCA} \label{Sec3}

Next, we review the supervised dynamic PCA (sdPCA) proposed by \citet{gao2023} for forecasting. 
The sdPCA is a factor-based forecasting approach, which took a major role in the literature of time series forecasting.
Tailored to incorporate dynamic information, the sdPCA has been shown to outperform some existing factor-based approaches such as the diffusion index model of \citet{SW2002, SW2002b}, and the scaled PCA of \citet{Huang2022}.

We first outline the sdPCA procedure. 
Given a forecast horizon $h$, the sdPCA first constructs intermediate predictions using each individual predictor and its lagged values.
For instance, one can regress $y_{t+h}$ on $x_{t,j}, x_{t-1,j}, \ldots, x_{t-q_{2}+1,j}$, where $q_{2} \in \mathbb{N}$ is a user-specified lag value, and obtain the fitted values
\begin{align} \label{3-1}
    \hat{\mu}_{j} + \sum_{k=0}^{q_{2}-1} \hat{\gamma}_{j,k}x_{t-k,j} := \hat{\mu}_{j} + \hat{x}_{t,j},
\end{align}
where $\hat{\mu}_{j}$ is the intercept estimate and $\hat{\gamma}_{0,1}, \ldots, \hat{\gamma}_{q_{2}-1,j}$ are the slope estimates. 
For different $j$'s, the number of lags $q_{2}$ used in the regression can differ. For instance, $q_{2}$ can be selected by the BIC.
Then, with the constructed intermediate predictions, $\hat{x}_{t,j}$'s, the sdPCA apply PCA to estimate a lower dimensional factor $\hat{\mathbf{g}}_{t} \in \mathbb{R}^{r}$, where $r < p$. 
Therefore, the data to which the PCA applies, $\hat{x}_{t,1}, \ldots, \hat{x}_{t,p}$, are in the same unit (as $y_{t}$), which scales the variables according to their predictive power.

Finally, one may employ a linear model for the predictive equation,
\begin{align} \label{3-2}
    y_{t+h} \sim \hat{\alpha} + \hat{\bm{\beta}}^{\top}\hat{\mathbf{g}}_{t},
\end{align}
where $\hat{\alpha}$ and $\hat{\bm{\beta}}$ are intercept and slope estimates respectively, and $\sim$ signifies that we run the linear regression while the underlying relationship of $y_{t+h}$ and $\hat{\mathbf{g}}_{t}$ may not be exactly linear. 
\citet{gao2023} also suggests to use Lasso to estimate \eqref{3-2} instead of the usual linear regression if the number of common 
factors is large.
Additionally, one can also include some lags of $\hat{\mathbf{g}}_{t}$ is \eqref{3-2} and let Lasso perform variable selection.

The sdPCA has several advantages over the conventional PCA methods.
First, the PCA is not scale-invariant. On the contrary, the sdPCA constructs predictors that are in the same unit, which naturally scales the predictors according to their predictive capabilities.
Second, instead of performing PCA directly on contemporaneous data $\mathbf{x}_{t}$, the sdPCA sources from the lagged information in $\mathbf{x}_{t-l}$, $l=0,1,\ldots, q_{2}-1$, where $\mathbf{x}_{t} = (x_{t,1}, \ldots, x_{t,p})^{\top}$.
For the conventional PCA to use the lagged information, one often needs to augment the data by appending the lagged variables so that PCA is performed on $(\mathbf{x}_{t}, \ldots, \mathbf{x}_{t-q_{2}+1})$, which leads to even higher dimensionality. 
Lastly, the usual PCA is performed in an unsupervised fashion, whereas the sdPCA constructs the factors in a supervised fashion.
While the usual principal component directions are not necessarily predictive of the response, factors extracted by sdPCA can potentially yield better forecasts.
In fact, \citet{gao2023} have showed that sdPCA has a lower mean square forecasting error than the approaches of \citet{SW2002, SW2002b} and \citet{Huang2022}.

In this paper, we employ the sdPCA to capitalize on the aforementioned properties.
However, for noisy high-dimensional data, the performance of sdPCA may be severely compromised.
Hence, careful dimension reduction before applying the sdPCA is desirable.
In the next section, we describe the proposed procedure, which combines GOGA and HDAIC with sdPCA to improve the accuracy of 
prediction. 

\section{The proposed GO-sdPCA} \label{Sec4}

In this section, we introduce the proposed method, GO-sdPCA, which screens variables by the GOGA and HDAIC and then estimates factors by the sdPCA approach.
In order to facilitate factor recovery, we apply a ``peeling'' technique to select more factor-relevant variables in the new 
procedure. 

To tackle the difficulties encountered by the factor-based approaches when applied to high-dimensional data, our method begins with dimension reduction by employing the GOGA introduced in Section \ref{Sec2}.
Because of the serial dependence in the data, it is beneficial to select variables based not only on cross-sectional correlation but also on the lagged information.
To this end, for each predictor $\mathbf{x}_{j} = (x_{1,j}, \ldots, x_{n,j})^{\top}$, we consider the group predictor,
\begin{align} \label{4-1}
    \mathbf{x}_{(j)} = \begin{pmatrix}
        x_{q_{1},j} & x_{q_{1} - 1, j} & \ldots & x_{1,j} \\
        x_{q_{1} + 1, j} & x_{q_{1}, j} & \ldots & x_{2,j} \\
        \vdots & \vdots & \ddots & \vdots \\
        x_{n-h, j} & x_{n-h-1,j} & \ldots & x_{n-h-q_{1}+1,j}
    \end{pmatrix}, \quad j=1,2,\ldots, p,
\end{align}
where $q_{1}$ is a pre-specified integer for the number of lagged values to consider, and $h \in \mathbb{N}$ is the forecast horizon. 
Let
\begin{align} \label{4-2}
    \mathbf{y} = (y_{q_{1}+h}, y_{q_{1}+h+1}, \ldots, y_{n})^{\top}
\end{align}
be the response vector.
Then, we employ the GOGA algorithm in Section \ref{Sec2}, with $\mathbf{x}_{(j)}$ and $\mathbf{y}$ substituted by \eqref{4-1} and \eqref{4-2}, respectively. 
Since each group predictor already incorporates past information, the GOGA algorithm can select variables while accounting for the dynamic information.

We also employ the HDAIC (defined in \eqref{2-2}) after GOGA to prevent the inclusion of irrelevant variables.
Since the above procedure depends on the number of GOGA iterations $K_{n}$, the HDAIC constant $C$ in \eqref{2-2}, the number of lags $q_{1}$ used in forming the group predictors, and the pool of candidate predictors $\mathcal{I} = \{1,2,\ldots,p\}$, we conveniently denote it as a set-valued function,
\begin{align}
    \hat{J} = \mathcal{A}(K_{n}, C, q_{1}, \mathcal{I}),
\end{align}
where $\hat{J}$ is the index set outputted by GOGA+HDAIC, as in \eqref{2-3}.
For completeness, we summarize the GOGA+HDAIC method introduced in Section \eqref{Sec2} in Algorithm \ref{alg:OGA+HDAIC}.

\begin{algorithm}[h!]
\DontPrintSemicolon
    \KwInput{Number of maximum iterations $K_{n}$, HDAIC parameter $C$, number of lags $q_{1}$, candidate set $\mathcal{I}$}
    \KwInit{$\mathbf{u}^{(0)} = \mathbf{y}$; selected index $\hat{J} = \emptyset$}
    \For{$k = 1,2, \ldots, K_{n}$}{
        Select 
        \begin{align*}
            \hat{j}_{k} &= \arg\min_{j \in \mathcal{I}} \Vert \mathbf{u}^{(k-1)} - \mathbf{x}_{(j)}(\mathbf{x}_{(j)}^{\top}\mathbf{x}_{(j)})^{-1} \mathbf{x}_{(j)}^{\top} \mathbf{u}^{(k-1)} \Vert^{2},
        \end{align*} 
        where $\mathbf{x}_{(j)}$ is defined in \eqref{4-1}. \\
        Update $\hat{J} \leftarrow \hat{J} \cup \{\hat{j}_{k}\}$ \\
        Update $\mathbf{u}^{(k)} = (\mathbf{I} - \mathbf{H}_{(k)})\mathbf{y}$,
        where $\mathbf{H}_{(k)}$ is the projection matrix associated with $\{\mathbf{x}_{(j)}: j \in \hat{J}\}$
    }
    Choose, as in \eqref{2-3},
    \begin{align*}
        \hat{m} = \arg\min_{1 \leq m \leq K} \mathrm{HDAIC}(\{\hat{j}_{1}, \ldots, \hat{j}_{m}\}),
    \end{align*}
    where HDAIC is defined in \eqref{2-2} with $C$ set to the inputted value.
    
    \KwOutput{selected indices $\{\hat{j}_{1}, \ldots, \hat{j}_{\hat{m}}\}$}
\caption{GOGA+HDAIC ($\mathcal{A}$)}
\label{alg:OGA+HDAIC}
\end{algorithm}

Because the orthogonal residuals are used in the greedy search, GOGA tends to differentiate highly correlated predictors and only selects those predictors that explain distinct (close to orthogonal) directions of the response. 
However, in factor estimation, the relevant variables loaded on the common factors tend to be highly correlated and failing to employ these correlated predictors may lose some statistical efficiency for the inference of the underlying factors (the blessing of dimensionality, \citealp{lam2011, gao2023}).
Therefore, to encourage GOGA to screen the factor-relevant, correlated predictors, we propose a technique, referred to as ``peeling,'' that allows careful inclusion of more variables. 
The idea is to repeatedly apply GOGA+HDAIC, with variables selected from previous implementations discarded from the candidate set. 
Hence, in each iteration, GOGA+HDAIC is forced to select a new set of variables that best predict $\mathbf{y}$. 
Formally, the peeling algorithm is described in Algorithm \ref{alg:peeling}.
Let $M$ be the number of peeling iterations. In the following, we denote the peeling procedure, which depends on $M$, $K_{n}$, and $C$, by the set-valued function $\mathcal{P}$. Namely,
\begin{align*}
    \hat{Q} = \mathcal{P}(M, K_{n}, C, q_{1}).
\end{align*}

\begin{algorithm}[h]
\DontPrintSemicolon
    \KwInput{Number of peeling iterations $M$, number of GOGA iterations $K_{n}$, HDAIC parameter $C$, number of lags $q_{1}$ in group predictors}
    \KwInit{$\hat{Q} = \emptyset$, $\mathcal{I} = \{1,2,\ldots,p\}$}
    \For{$m = 1,2, \ldots, M$}{
        Run GOGA+HDAIC
        \begin{align*}
            \hat{J}^{(m)} = \mathcal{A}(K_{n}, C, q_{1}, \mathcal{I})
        \end{align*} \\
        Update $\hat{Q} \leftarrow \hat{Q} \cup \hat{J}^{(m)}$ \\
        Discard selected variables from the candidate set $\mathcal{I} \leftarrow \mathcal{I} - \hat{Q}$
    }
    \KwOutput{selected indices $\hat{Q}$}
\caption{Peeling ($\mathcal{P}$)}
\label{alg:peeling}
\end{algorithm}

It is worth noting that peeling is different from simply running GOGA for many iterations. 
Although GOGA also selects distinct variables along its iteration because of orthogonalization (that is, GOGA does not revisit any previously selected variable), peeling, which discards previously selected variables from the candidate set, would produce very different results.
For high-dimensional data, running many iterations of GOGA will lead to extremely small residuals, which may no longer carry sufficient variations to detect the remaining relevant variables (with finite sample). 
On the other hand, peeling \textit{re-starts} GOGA in every iteration, using $\mathbf{y}$ instead of the previous residuals in the GOGA algorithm. 
Since GOGA is re-started with a smaller pool of candidate variables, the residuals in peeling will not shrink to zero after many (potentially more than $n$) variables are already contained in $\hat{Q}$.
We also remark that the idea of peeling is akin to the random forest which uses randomly selected variables in building each tree.
Hence, peeling can be used to detect the weak predictors in the case where a few of the predictors are highly predictive and many others are only weakly predictive \citep[see, e.g. Chpt. 8 of][]{james2021}.

After variable selection by peeling, sdPCA is employed to estimate the factors $\hat{\mathbf{f}}_{t}$ from the selected variables. 
Let $q_{2}$ be the number of lags used in constructing the intermediate predictions \eqref{3-1}.
Finally, the predictive model
\begin{align} \label{4-3}
    y_{t+h} = \sum_{k=1}^{q_{3}} \alpha_{k} y_{t-q+1} + \bm{\beta}^{\top}\hat{\mathbf{f}}_{t} + \epsilon_{t+h},
\end{align}
is estimated by OLS or Lasso, where $q_{3}$ is the number of autoregressive variables.
The above procedure, which combines GOGA, peeling, and sdPCA, is called GO-sdPCA.

In closing this section, we briefly discuss the selection of the number of lags. 
In practice, the number of lags $q_{1}$ used in GOGA, $q_{2}$ used in the sdPCA step, and $q_{3}$ in the predictive equation can all differ. 
If $q_{1}$ is large, we can detect the predictors whose distant lags are predictive of the response.
However, each step of GOGA iteration will then consume more useful variations, so GOGA will consequently select a smaller number of grouped predictors, which can be counterproductive for the subsequent factor estimation.
Thus it is advisable to use a smaller $q_{1}$ in the screening step, which leads to more predictors selected, and use a larger $q_{2}$ in the sdPCA step to extract information from the past.

\section{Simulation studies} \label{Sec5}
In this section, we assess the finite-sample performance of the proposed GO-sdPCA method via simulation studies.
Some existing factor-based methods are employed as benchmarks, such as the diffusion index approach of \citet{SW2002, SW2002b}, the time series factor model of \citet{lam2011}, and the supervised dynamic PCA of \citet{gao2023}. These benchmarks are referred to as SW, LYB, and sdPCA, respectively.

After factor estimation, the predictive model
\begin{align} \label{5-1}
    y_{t+1} = \sum_{k=1}^{q} \alpha_{k} y_{t-q+1} + \bm{\beta}^{\top}\hat{\mathbf{f}}_{t} + \epsilon_{t+1},
\end{align}
is estimated by OLS, where $q$ is an integer specified later, and $\hat{\mathbf{f}}_{t}$ is constructed using different approaches.
For the proposed GO-sdPCA (shorthanded as GsP$^{*}$ hereafter), we set $C = 2$ in the HDAIC definition \eqref{2-2} and use 10 peeling iterations.
In each peeling iteration, GOGA is applied with $q_{1}$ set to 2 for the group predictors.
Then, in the sdPCA step, $r$ factors are extracted with $q_{2} = q$ lags used in constructing the intermediate predictions in \eqref{3-1}.
To demonstrate the usefulness of the peeling technique, we also consider implementing GO-sdPCA by naively combining GOGA+HDAIC with sdPCA.
That is, we only run one peeling iteration and the variables selected are exactly the ones selected by GOGA+HDAIC.
This method is denoted as GsP in the sequel. 
Similarly, the forecasts of sdPCA are constructed by estimating \eqref{5-1} with the factors estimated as in Section \ref{Sec3}. 
The time series factors of \citet{lam2011} are estimated using the eigenanalysis of a non-negative definite matrix computed from the autocovariance matrices at nonzero lags. 
In our implementation, $q$ lags of past predictors are used by LYB in the eigenanalysis.
Finally, SW follows that of \citet{SW2002, SW2002b} and uses PCA to extract the factors. 

In addition, we employ some commonly used alternatives, including the Lasso of \citet{tib1996} and the random forests of \citet{Breiman2001} as competing methods.
The Lasso is a versatile tool for building sparse linear regression models, while the random forest (RF) excels in capturing non-linear relationships.
Recently, \citet{Chi2022} investigated the asymptotic consistency of RF for high-dimensional data.
See also \citet{Saha2023} for application of RF to dependent data.
The tuning parameters for the Lasso are selected by the BIC as suggested by \citet{medeiros2016}, whereas we adopt the hyper-parameters for RF as recommended by the \texttt{randomForest} \citep{randomForest} package in \texttt{R}.
Therefore, about one-third of the candidate variables is randomly selected at each split.
For both methods, $q$ lags of the dependent variable and the predictors are used for fair comparison. 

\subsection{Simulation designs and results} 

In the simulations, we consider three data generating processes (DGP) to generate the synthetic data.
Throughout the experiments, we use one-step-ahead forecasts ($h=1$) where each method makes a forecast for $y_{n+1}$, which is not in the training sample.
The root mean squared forecast errors, averaged over 500 Monte-Carlo simulations, are used for comparing different approaches.

\begin{dgp}
    Let $\mathbf{f}_{t} \sim_{\mathrm{i.i.d.}} N(0, \mathbf{I}_{r_{\mathrm{DGP}}})$, where $r_{\mathrm{DGP}} \in \mathbb{N}$ is the number of underlying factors. The predictors $\mathbf{x}_{t} \in \mathbb{R}^{p}$ are generated by
    \begin{align*}
        \mathbf{x}_{t} = \mathbf{B}\mathbf{f}_{t} + 2\bm{\delta}_{t},
    \end{align*}
    where $\{\bm{\delta}_{t}\}$ are independent $p$-dimensional $t$-distributed random vectors with independent components and five degrees of freedom, and $\mathbf{B} \in \mathbb{R}^{p \times r_{\mathrm{DGP}}}$ has independent $\mathrm{Unif}(-2,2)$ entries, with $p-s$ rows randomly set to zero. That is, $\mathbf{B}$ only has $s$ nonzero rows. 
    Randomly generate $\bm{\beta}_{1} = (\beta_{1,1}, \ldots, \beta_{r_{\mathrm{DGP}},1})^{\top}$ and $\bm{\beta}_{2} = (\beta_{1,2}, \ldots, \beta_{r_{\mathrm{DGP}}, 2})^{\top}$ via $\beta_{j,1} \sim \mathrm{Unif}(1.0, 2.5)$ and $\beta_{j,2} \sim \mathrm{Unif}(-2.0, -0.8)$. 
    Finally,
    \begin{align*}
        y_{t} = 0.6y_{t-1} + 0.2 y_{t-2} + \bm{\beta}_{1}^{\top} \mathbf{f}_{t-1} + \bm{\beta}_{2}^{\top} \mathbf{f}_{t-2} + \epsilon_{t},
    \end{align*}
    where $\{\epsilon_{t}\}$ are independent standard Gaussian. 
\end{dgp}

In this DGP, the parameter $s$ dictates both the number of relevant variables and the strength in recovering the factors.
The larger $s$ is, the stronger the factors are.
Factor strength plays a critical role in factor recovery \citep{SW2002, SW2002b, lam2011}.
In practice, $s$ is seldom known.
Therefore, we will consider the cases $s \in \{0.25n, 0.5n, 0.75n\}$, where $n$ is the sample size, to see whether the methods adapt well to various levels of factor strength. 

Table \ref{tab:DGP1} reports the root mean squared forecast error (RMSFE) averaged over 500 Monte Carlo simulations.
Across all sparsity levels $s \in \{0.25n, 0.5n, 0.75n\}$, the proposed GsP$^{*}$ delivered the most accurate forecasts and 
the sdPCA ranks the second. This suggests that the proposed peeling 
procedure and GOGA improves accuracy in forecasting. 
The GsP, which naively combines GOGA+HDAIC with sdPCA, shows limited forecasting capabilities with RMSFE  being higher 
than those of the Lasso. This indicates, again, the peeling strategy markedly improved the forecasting performance by selecting more factor-relevant variables.
The other forecasting methods, including SW, LYB and RF, 
seem to suffer from the effect of employing many irrelevant variables in the high-dimensional data. We remark that DGP 1 is essentially 
a sparse model because $r_{DGP}$ used is relatively small. 
Therefore, it is not surprising to 
see that Lasso fares reasonably well. 

\begin{table}[h]
\centering
\caption{Root mean squared forecast error of various competing methods. The data are generated from DGP 1 and the results are averaged over 500 Monte Carlo simulations. $n$ and $p$ stand for the sample size and number of observed predictors. $r_{\mathrm{DGP}}$ is defined in DGP 1, and $r$ is the number of factors extracted using various methods.}
\label{tab:DGP1}
\begin{tabular}{@{}rrrrrrrr@{}}
\toprule
& GsP$^{*}$ & GsP & sdPCA & SW & LYB & Lasso & RF \\ 
$(r_\mathrm{DGP}, s)$ & \multicolumn{7}{c}{$(n, p, r) = (200, 1000, 10)$} \\ \midrule
(5, 50) & 1.894 & 2.443 & 1.993 & 2.657 & 2.136 & 2.207 & 2.860 \\
(10, 100) & 2.406 & 3.299 & 2.488 & 3.288 & 2.909 & 2.831 & 4.353 \\
(15, 150) & 3.315 & 3.908 & 3.584 & 5.368 & 5.201 & 3.537 & 5.776 \\
\multicolumn{8}{c}{$(n, p, r) = (400, 2000, 30)$} \\ \midrule
(10, 100) & 2.391 & 2.947 & 2.590 & 3.482 & 2.077 & 2.544 & 4.428 \\
(20, 200) & 3.020 & 4.355 & 3.242 & 4.626 & 3.657 & 3.679 & 6.604 \\ 
(30, 300) & 3.818 & 5.581 & 4.155 & 5.666 & 4.651 & 4.450 & 8.253 \\\bottomrule
\end{tabular}
\end{table}

\begin{dgp}
    In this DGP, $\mathbf{x}_{t}$ is generated via a vector MA(1) model:
    \begin{align*}
        \mathbf{x}_{t} = \bm{\delta}_{t} + 0.8 \mathbf{B} \bm{\delta}_{t-1},
    \end{align*}
    where $\mathbf{B}$ is a randomly drawn $p \times p$ matrix of rank $r_{\mathrm{DGP}}$. 
    The coefficients $\bm{\beta}_{1} = (\beta_{1,1}, \ldots, \beta_{p,1})^{\top}$ and $\bm{\beta}_{2} = (\beta_{1,2}, \ldots, \beta_{p, 2})^{\top}$ are randomly generated via 
    $\beta_{j,1} \sim \mathrm{U}(1.0, 3.0)$ and $\beta_{j,2} \sim \mathrm{U}(-2.5, -0.5)$. But, for 
    a set of random indices $J$ with cardinality equal to $s$ (i.e., $\sharp(J) = s$), we set $\beta_{k,1} = \beta_{k,2} = 0$ for $k \notin J$. In this way, $\bm{\beta}_{1}$ and $\bm{\beta}_{2}$ share the same nonzero coordinates. 
    Finally,
    \begin{align*}
        y_{t} = 0.6 y_{t-1} + 0.2 y_{t-2} + \bm{\beta}_{1}^{\top} \mathbf{x}_{t-1} + \bm{\beta}_{2}^{\top} \mathbf{x}_{t-2} + \epsilon_{t}.
    \end{align*}
\end{dgp}

In this example, the relevant predictors have a direct impact on the response, instead of through any common factors.
Additionally, when $s$ is large, it is very difficult to recover the regression coefficients because of the lack of sparsity.
Therefore, DGP 2 fits neither the factor model nor the sparse linear model frameworks.
Nevertheless, the covariance matrix of $\mathbf{x}_{t}$ has a special structure. Observe that 
\begin{align*}
    \mathbb{E}(\mathbf{x}_{t}\mathbf{x}_{t}^{\top}) = \mathbf{I}_{p} + 0.64 \mathbf{B}\mathbf{B}^{\top}.
\end{align*}
That is, the covariance matrix of $\mathbf{x}_{t}$ has the form of a spiked matrix. 


Table \ref{tab:DGP2} reports the RMSFEs of the competing methods considered.
The sdPCA fares the best as it yields the smallest RMSFE, 
especially when $s$ is small.
This reflects that the sdPCA can better utilize the spiked covariance structure by selecting contributions from relevant predictors.
However, the performance of GsP$^{*}$ and sdPCA converge when $s = 0.75n$, and both of them compare favorably against the other alternatives. As expected, for non-sparse models, Lasso and random 
forest do not work well. 

\begin{table}[h]
\centering
\caption{Root mean squared forecast error of the competing methods, divided by 1000. Data are generated from DGP 2 and the results are averaged over 500 Monte Carlo simulations. $n$ and $p$ stand for the sample size and number of observed predictors. $r_{\mathrm{DGP}}$ is defined in DGP 2, $r$ is the number of factors extracted using various methods, and $s$ is a sparsity parameter.}
\label{tab:DGP2}
\begin{tabular}{@{}rrrrrrrr@{}}
\toprule
& GsP$^{*}$ & GsP & sdPCA & SW & LYB & Lasso & RF \\ 
$(r_\mathrm{DGP}, s)$ & \multicolumn{7}{c}{$(n, p, r) = (200, 1000, 10)$} \\ \midrule
(5, 50) & 0.857 & 0.883 & 0.695 & 1.017 & 0.966 & 0.870 & 2.017 \\
(10, 100) & 2.833 & 3.137 & 2.834 & 3.436 & 3.436 & 3.802 & 6.971 \\
(15, 150) & 6.559 & 7.068 & 6.379 & 7.801 & 7.761 & 9.532 & 16.264 \\
\multicolumn{8}{c}{$(n, p, r) = (400, 2000, 30)$} \\ \midrule
(10, 100) & 4.091 & 4.547 & 2.259 & 5.224 & 4.917 & 5.118 & 8.977 \\
(20, 200) & 15.891 & 17.357 & 8.411 & 20.012 & 19.411 & 20.961 & 35.132 \\ 
(30, 300) & 34.557 & 37.956 & 34.543 & 42.904 & 42.905 & 50.996 & 70.180 \\\bottomrule
\end{tabular}
\end{table}

\begin{dgp}
    In this example, the predictors follow a VAR model:
    \begin{align*}
        \mathbf{x}_{t} = \mathbf{B}\mathbf{x}_{t-1} + \bm{\delta}_{t},
    \end{align*}
    where $\{\bm{\delta}_{t}\}$ are independent $p$-dimensional standard Gaussian processes. Let $\tilde{\mathbf{B}}$ be a randomly generated $p \times p$ matrix of rank $r_{\mathrm{DGP}}$.
    Then the AR coefficient matrix is constructed as
    \begin{align*}
        \mathbf{B} = \frac{\tilde{\mathbf{B}}}{1.05 \Vert \tilde{\mathbf{B}} \Vert},
    \end{align*}
    where $\Vert \cdot \Vert$ denotes the operator norm. The target variable is generated via
    \begin{align*}
        y_{t} = 0.5 y_{t-1} + \bm{\beta}^{\top}\mathbf{x}_{t-1} + \epsilon_{t},
    \end{align*}
    where $\{\epsilon_{t}\}$ are independent standard Gaussian 
    variates and $\bm{\beta} = (\beta_{1}, \ldots, \beta_{p})^{\top}$ with 
    \begin{align*}
        \beta_{j} = \left\{ \begin{array}{cc}
           (-1)^{j}u_{j},  & 1 \leq j \leq s  \\
            0, & \mathrm{otherwise}
        \end{array}\right.,
    \end{align*}
    in which $u_{j} \sim_{\mathrm{i.i.d.}} \mathrm{U}(0.1, 3.0)$.
\end{dgp}

In this DGP, the predictors, following a high-dimensional VAR(1) model, exhibit complex dynamics and correlations.
Without simplifying structures such as latent factors, spiked covariance matrices, and sparsity, forecasting becomes difficult.
Table \ref{tab:DGP3} reports the RMSFEs of various competing methods.
With small $s$, which corresponds to sparse models, Lasso can suitably choose the predictors 
and yield relatively accurate forecasts. 
However, as $s$ increases, its RMSFE quickly increases and becomes similar to those of the factor-based approaches. The factor-based approaches as well as the RF performed similarly under this DGP. 
This example also demonstrates that the sdPCA method may encounter loss in prediction accuracy 
if the number of selected common factors is under-specified; see the case of $r_\mathrm{GDP} = 15$ and $r=10$. 

Overall, our simulation studies show that no forecasting method always dominates the competing methods 
used in the study, but the proposed GsP$^*$ procedure can be effective in some cases.

\begin{table}[h]
\centering
\caption{Root mean squared forecast error of various competing methods. Data are generated from DGP 3 and the results are averaged over 500 Monte Carlo simulations. $n$ and $p$ stand for the sample size and number of observed predictors. $r_{\mathrm{DGP}}$ is defined in DGP 3, and $r$ is the number of factors extracted using various methods.}
\label{tab:DGP3}
\begin{tabular}{@{}rrrrrrrr@{}}
\toprule
& GsP$^{*}$ & GsP & sdPCA & SW & LYB & Lasso & RF \\ 
$(r_\mathrm{DGP}, s)$ & \multicolumn{7}{c}{$(n, p, r) = (200, 1000, 10)$} \\ \midrule
(5, 50) & 12.255 & 12.166 & 12.347 & 12.565 & 12.469 & 7.537 & 12.813 \\
(10, 100) & 17.818 & 18.857 & 17.608 & 17.371 & 17.701 & 16.670 & 18.427 \\
(15, 150) & 23.142 & 24.417 & 23.041 & 21.777 & 22.027 & 21.125 & 22.660 \\
\multicolumn{8}{c}{$(n, p, r) = (400, 1000, 30)$} \\ \midrule
(10, 100) & 16.523 & 17.540 & 17.767 & 18.647 & 18.968 & 14.833 & 18.602 \\
(20, 200) & 27.266 & 29.110 & 26.063 & 25.291 & 25.717 & 25.953 & 26.503 \\ 
(30, 300) & 33.763 & 35.635 & 32.438 & 31.672 & 31.827 & 31.340 & 32.427 \\\bottomrule
\end{tabular}
\end{table}

\section{Empirical examples} \label{Sec6}

In this section, we apply the proposed method to two real data sets. The first data set is the U.S. macroeconomic data and the other consists of Taiwan particulate matter (PM$_{2.5}$) measurements.
Forecasting plays an essential role in the applications pertaining to these two data sets, despite that both of them contain high-dimensional predictors. 
Moreover, these two datasets have different characteristics, which enable us to examine the forecasting performance of various forecasting methods available in the literature.
In addition to the proposed GO-sdPCA approach, the competing methods  used in the simulation studies, such as sdPCA, SW, LYB, Lasso, and RF, are also employed in this section as benchmarks.

For both datasets, we consider the rolling-window $h$-step-ahead forecasting. 
Let $\{y_{t}\}$ be the target variable of interest. 
For predicting $y_{t+h}$, the factor-based methods use the predictive equation
\begin{align*}
    \hat{y}_{t+h} = \sum_{k=1}^{q} \alpha_{k} y_{t-q+1} + \bm{\beta}^{\top}\hat{\mathbf{f}}_{t} + \epsilon_{t+h},
\end{align*} 
where $\hat{\mathbf{f}}_{t}$ is the vector of estimated common factors by different methods. 
For Lasso and RF, $y_{t}, y_{t-1} \ldots, y_{t-q+1}, \mathbf{x}_{t}, \ldots, \mathbf{x}_{t-q+1}$ are used as potential predictors.

\subsection{U.S. macroeconomic data}
\begin{figure} 
    \centering
    \begin{subfigure}[t]{0.49\textwidth}
        \includegraphics[width=\linewidth]{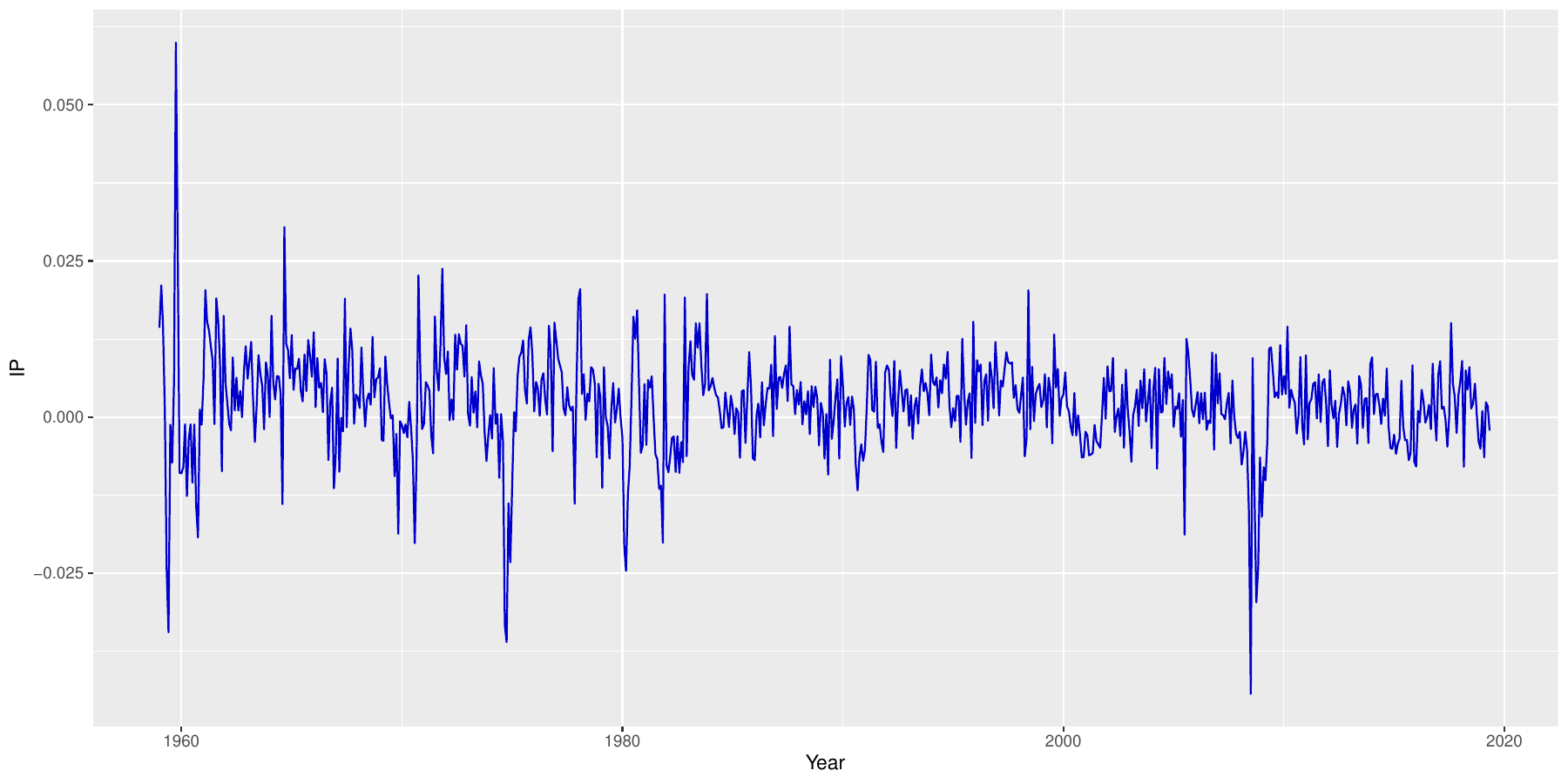}
        \caption{Industrial production}
    \end{subfigure}
    \begin{subfigure}[t]{0.49\textwidth}
        \includegraphics[width=\linewidth]{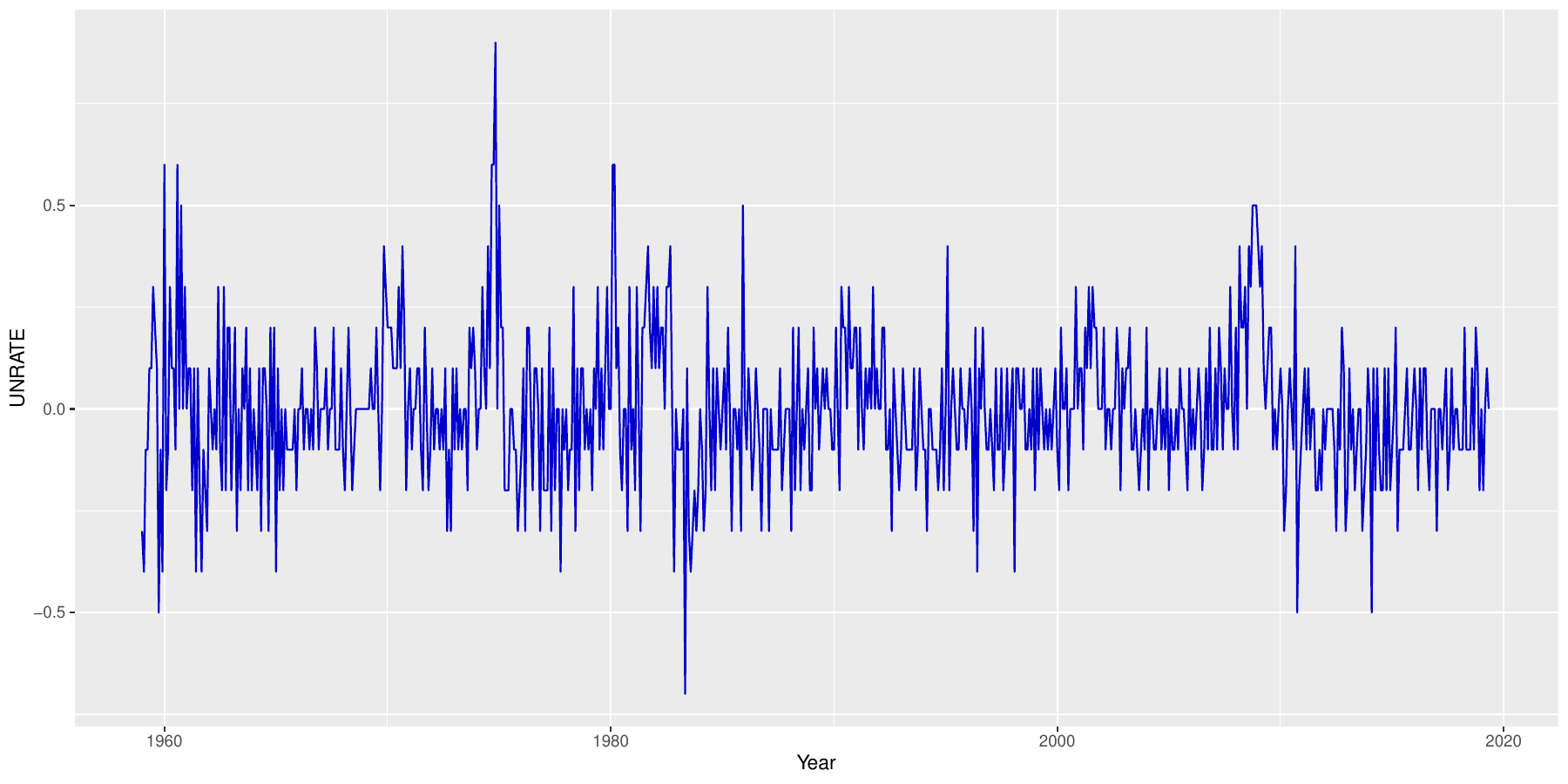}
        \caption{Unemployment rate}
    \end{subfigure} \\
    \begin{subfigure}[t]{0.49\textwidth}
        \includegraphics[width=\linewidth]{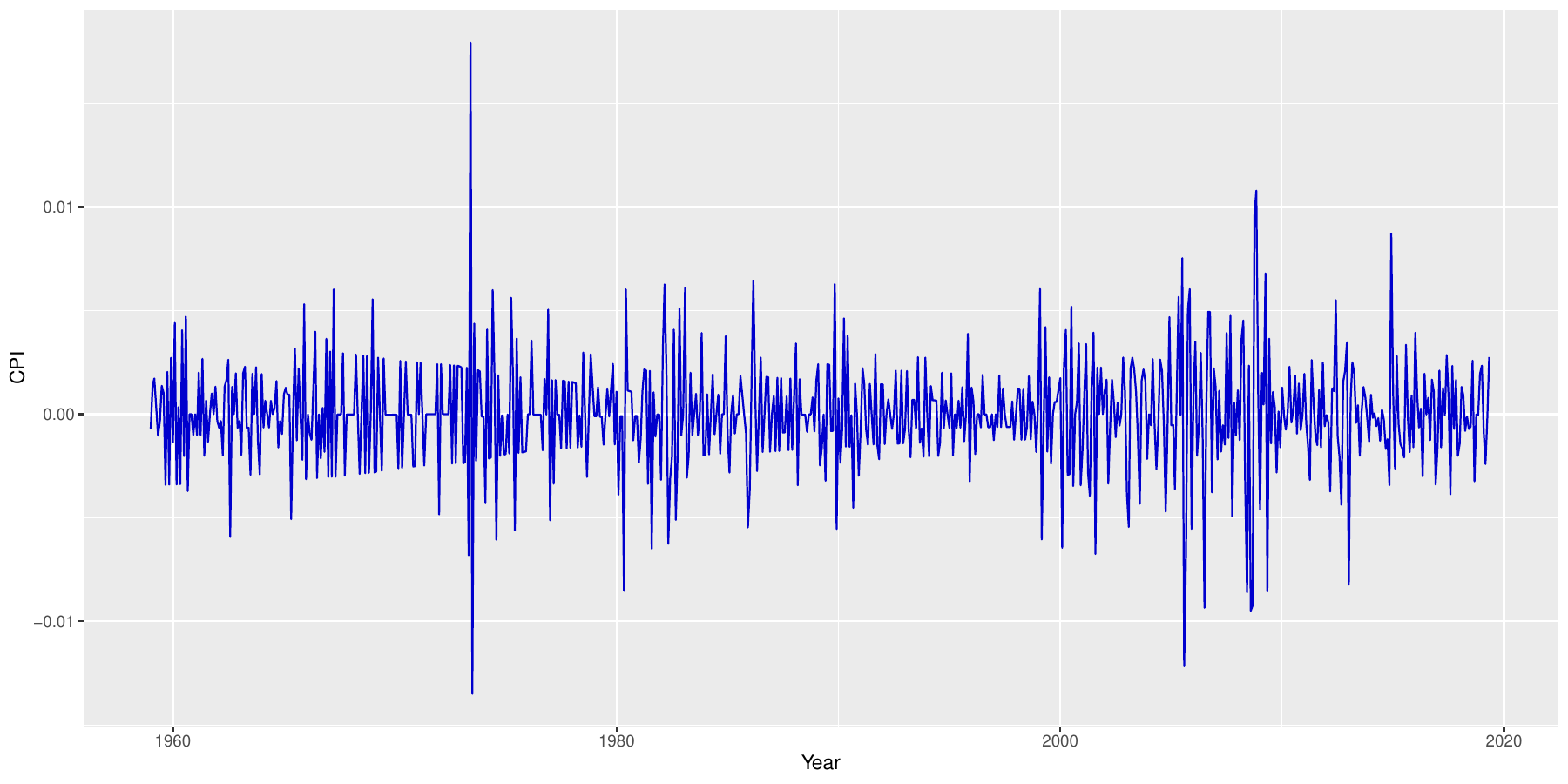}
        \caption{CPI}
    \end{subfigure}
    \begin{subfigure}[t]{0.49\textwidth}
        \includegraphics[width=\linewidth]{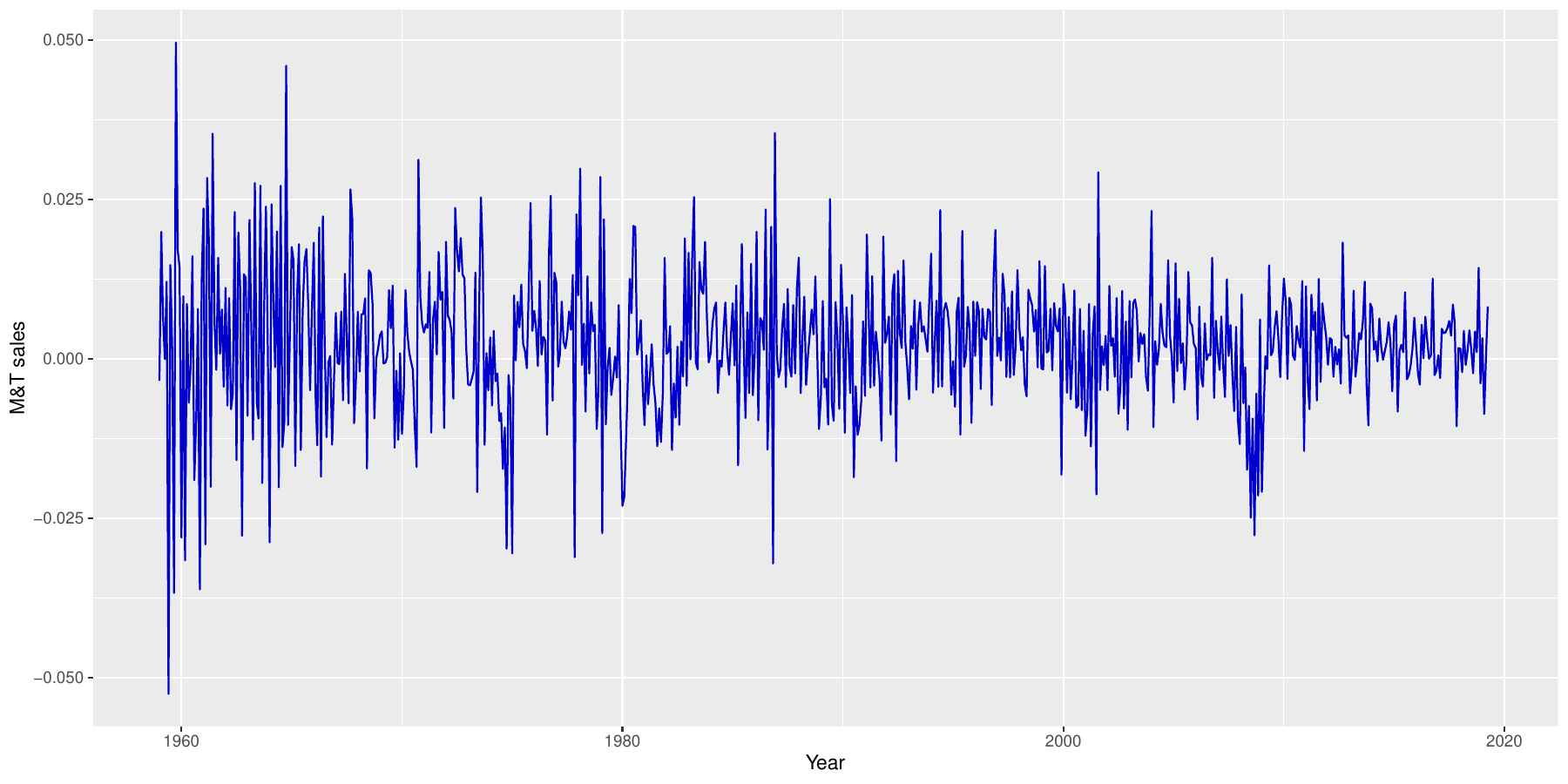}
        \caption{M\&T sales}
    \end{subfigure}
    \caption{Time plots of the (transformed) macroeconomic time series of interest.}
    \label{fig:1}
\end{figure}

First, we consider the U.S. monthly macroeconomic data from January 1973 to June 2019.
The data are from the FRED-MD database maintained by St. Louis Federal Reserve at \url{https://research.stlouisfed.org/econ/mccracken/fred-databases/}. 
We transform the time series to stationarity according to \citet{maccraken2016}, and, after discarding some variables containing missing values, there are 125 macroeconomic time series. 
Among them, we focus on predicting (1) industrial production index, (2) unemployment rate, (3)  CPI-All, and (4) Real manufacturing and trade industries sales (M\&T sales). Time plots of these four target series after transformations are depicted in Figure \ref{fig:1}. 

For this data set, we consider $h = 1$ and different combinations of $q$ and $r$ (the number of factors extracted). 
Since $q$ lags of the predictors are used, there are $125q$ total predictors, which exceeds the sample size in each window.
The last twenty years of data (240 time periods) are reserved for testing. 
In addition to the root mean squared forecasting errors, we also report the $p$-values of the Diebold-Mariano test \citep{Diebold1995, HARVEY1997} against the alternative hypothesis that the proposed GsP$^{*}$ procedure is more accurate.

Tables \ref{tab:IP}--\ref{tab:MTsales} present the results.
The proposed GO-sdPCA achieved the lowest RMSFE for three of the targeted series: industrial production index, unemployment rate, and CPI. 
It notably outperformed the factor-based alternatives sdPCA, SW, and LYB with high confidence in forecasting these three series. 
In addition, no other methods demonstrated such consistent effectiveness across the targeted series.


\begin{table}[h]
\centering
\caption{Root mean squared forecast errors $\times 100$ for predicting industrial production index. The lowest RMSFE achieved by each method is in boldface. Among these boldfaced values, the lowest two values are marked in red. $q$ denotes the number of lags used in estimation and $r$ is the number of factors extracted. GsP$^*$ denotes the proposed GO-sdPCA. The figures in the parentheses are the $p$-values of the Diebold-Mariano test of whether GsP$^{*}$ is more accurate.}
\begin{tabular}{@{}cccccccccc@{}}
\toprule
      \multicolumn{10}{c}{IP (RMSE {\footnotesize $\times 100$}); $h = 1$}       \\ 
$q$     & \multicolumn{3}{c}{2} & \multicolumn{3}{c}{3} & \multicolumn{3}{c}{4} \\ \cmidrule(lr){2-4} \cmidrule(lr){5-7} \cmidrule(lr){8-10}
Lasso &  \multicolumn{3}{c}{0.626} & \multicolumn{3}{c}{\bf 0.625} & \multicolumn{3}{c}{0.628} \\
      &  \multicolumn{3}{c}{} & \multicolumn{3}{c}{(0.053)} & \multicolumn{3}{c}{} \\
RF    &  \multicolumn{3}{c}{0.621} & \multicolumn{3}{c}{\bf 0.614} & \multicolumn{3}{c}{0.619} \\
      &  \multicolumn{3}{c}{} & \multicolumn{3}{c}{(0.046)} & \multicolumn{3}{c}{} \\
$r$     & 2     & 4     & 6     & 2     & 4     & 6     & 2     & 4     & 6     \\ \cmidrule(lr){2-4} \cmidrule(lr){5-7} \cmidrule(lr){8-10}
GsP$^*$  &  0.603  & 0.590  & 0.585   & 0.580  & 0.578 & 0.570  & 0.574  & \color{red}{\bf 0.570}  & 0.571      \\
sdPCA &  0.663  & 0.792  & 0.727   & 0.603  & 0.676 & 0.683  & \color{red}{\bf 0.601}  & 0.832  & 0.868      \\
& & & & & & & (0.024) & \\
SW    &  0.631  & 0.636  & 0.640   & 0.623  & 0.629 & 0.634  & {\bf 0.617}  & 0.624  & 0.629      \\
& & & & & & & (0.017) & \\
LYB   &  0.631  & 0.634  & 0.636   & 0.622  & 0.621 & 0.623  & 0.615 & {\bf 0.609} & 0.612 \\
& & & & & & & &  (0.037) & \\
\bottomrule
\end{tabular}
\label{tab:IP}
\end{table}

\begin{table}[h]
\centering
\caption{Root mean squared forecast errors for predicting unemployment rate. The lowest RMSFE achieved by each method is in boldface. Among these boldfaced values, the lowest two values are marked in red. $q$ denotes the number of lags used in estimation and $r$ is the number of factors extracted. GsP$^*$ denotes the proposed GO-sdPCA. The figures in the parentheses are the $p$-values of the Diebold-Mariano test of whether GsP$^{*}$ is more accurate.}
\begin{tabular}{@{}cccccccccc@{}}
\toprule
      \multicolumn{10}{c}{UNRATE (RMSE); $h = 1$}       \\ 
$q$     & \multicolumn{3}{c}{2} & \multicolumn{3}{c}{3} & \multicolumn{3}{c}{4} \\ \cmidrule(lr){2-4} \cmidrule(lr){5-7} \cmidrule(lr){8-10}
Lasso &  \multicolumn{3}{c}{0.138} & \multicolumn{3}{c}{\bf 0.138} & \multicolumn{3}{c}{0.139} \\
 &  \multicolumn{3}{c}{} & \multicolumn{3}{c}{(0.108)} & \multicolumn{3}{c}{} \\
RF    &  \multicolumn{3}{c}{0.135} & \multicolumn{3}{c}{\color{red}{\bf 0.134}} & \multicolumn{3}{c}{0.135} \\
 &  \multicolumn{3}{c}{} & \multicolumn{3}{c}{(0.280)} & \multicolumn{3}{c}{} \\
$r$     & 2     & 4     & 6     & 2     & 4     & 6     & 2     & 4     & 6     \\ \cmidrule(lr){2-4} \cmidrule(lr){5-7} \cmidrule(lr){8-10}
GsP$^*$  &  0.134  & 0.135  & 0.133   & 0.133  & 0.134 & 0.132  & 0.133  & 0.134  & \color{red}{\bf 0.132}      \\
sdPCA &  0.182  & 0.156  & 0.136   & 0.183  & {\bf 0.135}  & 0.144  & 0.229  & 0.181  & 0.146      \\
& & & & & (0.080) & & & & \\
SW    &  0.147  & 0.145  & 0.144   & 0.146  & {\bf 0.144} & 0.144  & 0.147  & 0.145  & 0.145      \\
& & & & & (0.002) & & & & \\
LYB   & 0.146   & {\bf 0.145}  & 0.152  & 0.145 & 0.146 & 0.147 & 0.147 & 0.146 & 0.149 \\
& & (0.008) & & & & & & & \\
\bottomrule
\end{tabular}
\label{tab:UNRATE}
\end{table}

\begin{table}[h]
\centering
\caption{Root mean squared forecast errors $\times 100$ for predicting CPI. The lowest RMSFE achieved by each method is in boldface. Among these boldfaced values, the lowest two values are marked in red. $q$ denotes the number of lags used in estimation and $r$ is the number of factors extracted. GsP$^*$ denotes the proposed GO-sdPCA. The figures in the parentheses are the $p$-values of the Diebold-Mariano test of whether GsP$^{*}$ is more accurate.}
\begin{tabular}{@{}cccccccccc@{}}
\toprule
      \multicolumn{10}{c}{CPI (RMSE {\footnotesize $\times 100$}); $h = 1$}       \\ 
$q$     & \multicolumn{3}{c}{2} & \multicolumn{3}{c}{3} & \multicolumn{3}{c}{4} \\ \cmidrule(lr){2-4} \cmidrule(lr){5-7} \cmidrule(lr){8-10}
Lasso &  \multicolumn{3}{c}{\color{red}{\bf 0.278}} & \multicolumn{3}{c}{0.281} & \multicolumn{3}{c}{0.282} \\
&  \multicolumn{3}{c}{(0.162)} & \multicolumn{3}{c}{} & \multicolumn{3}{c}{} \\
RF    &  \multicolumn{3}{c}{\bf 0.302} & \multicolumn{3}{c}{0.303} & \multicolumn{3}{c}{0.304} \\
&  \multicolumn{3}{c}{(0.014)} & \multicolumn{3}{c}{} & \multicolumn{3}{c}{} \\
$r$     & 2     & 4     & 6     & 2     & 4     & 6     & 2     & 4     & 6     \\ \cmidrule(lr){2-4} \cmidrule(lr){5-7} \cmidrule(lr){8-10}
GsP$^*$  &  0.286  & 0.281  & 0.273   & 0.277  & 0.270 & \color{red}{\bf 0.268}  & 0.284  & 0.277  & 0.269      \\
sdPCA &  0.334  & {\bf 0.287}  & 0.319   & 0.301  & 0.366  & 0.368  & 0.506  & 0.464  & 0.304      \\
& & (0.039) & & & & & & & \\
SW    &  0.300  & 0.301  & 0.294   & 0.291  & 0.292  & {\bf 0.282}  & 0.294  & 0.293  & 0.283      \\
& & & & & & (0.079) & & & \\
LYB   &  0.300  & 0.300  & 0.304   & {\bf 0.293}  & 0.294  & 0.296  & 0.293  & 0.294  & 0.299 \\
& & & & (0.059) & & & & & \\
\bottomrule
\end{tabular}
\label{tab:CPI}
\end{table}

\begin{table}[h]
\centering
\caption{Root mean squared forecast errors $\times 100$ for predicting MT sales. The lowest RMSFE achieved by each method is in boldface. Among these boldfaced values, the lowest two values are marked in red. $q$ denotes the number of lags used in estimation and $r$ is the number of factors extracted. GsP$^*$ denotes the proposed GO-sdPCA. The figures in the parentheses are the $p$-values of the Diebold-Mariano test of whether GsP$^{*}$ is more accurate.}
\begin{tabular}{@{}cccccccccc@{}}
\toprule
      \multicolumn{10}{c}{MT sales (RMSE {\footnotesize $\times 100$}); $h = 1$}       \\ \midrule
$q$     & \multicolumn{3}{c}{2} & \multicolumn{3}{c}{3} & \multicolumn{3}{c}{4} \\ \cmidrule(lr){2-4} \cmidrule(lr){5-7} \cmidrule(lr){8-10}
Lasso &  \multicolumn{3}{c}{0.783} & \multicolumn{3}{c}{\bf 0.773} & \multicolumn{3}{c}{0.778} \\
&  \multicolumn{3}{c}{} & \multicolumn{3}{c}{(0.203)} & \multicolumn{3}{c}{} \\
RF    &  \multicolumn{3}{c}{\color{red}{\bf 0.746}} & \multicolumn{3}{c}{0.748} & \multicolumn{3}{c}{0.758} \\
&  \multicolumn{3}{c}{(0.561)} & \multicolumn{3}{c}{} & \multicolumn{3}{c}{} \\
$r$     & 2     & 4     & 6     & 2     & 4     & 6     & 2     & 4     & 6     \\ \cmidrule(lr){2-4} \cmidrule(lr){5-7} \cmidrule(lr){8-10}
GsP$^*$  &  0.777  & 0.806  & 0.815   & {\bf 0.750}  & 0.775 & 0.775  & 0.759  & 0.784  & 0.801      \\
sdPCA &  2.104  & 2.092  & 2.100   & {\bf 1.893}  & 1.952  & 1.962  & 2.166  & 2.270  & 2.156      \\
& & & & (0.113) & & & & & \\
SW    &  0.761  & 0.745  & 0.774   & 0.750  & \color{red}{\bf 0.737}  & 0.764  & 0.758  & 0.745  & 0.772      \\
& & & & & (0.736) & & & & \\
LYB   &  0.796  & 0.792  & 0.799   & {\bf 0.768}  & 0.784  & 0.786  & 0.772  & 0.773  & 0.797 \\
& & & & (0.215) & & & & & \\
\bottomrule
\end{tabular}
\label{tab:MTsales}
\end{table}

\subsection{Particulate matters in Taiwan}
\begin{figure}[h!]
    \centering
    \begin{subfigure}[t]{0.49\textwidth}
        \includegraphics[width=\linewidth]{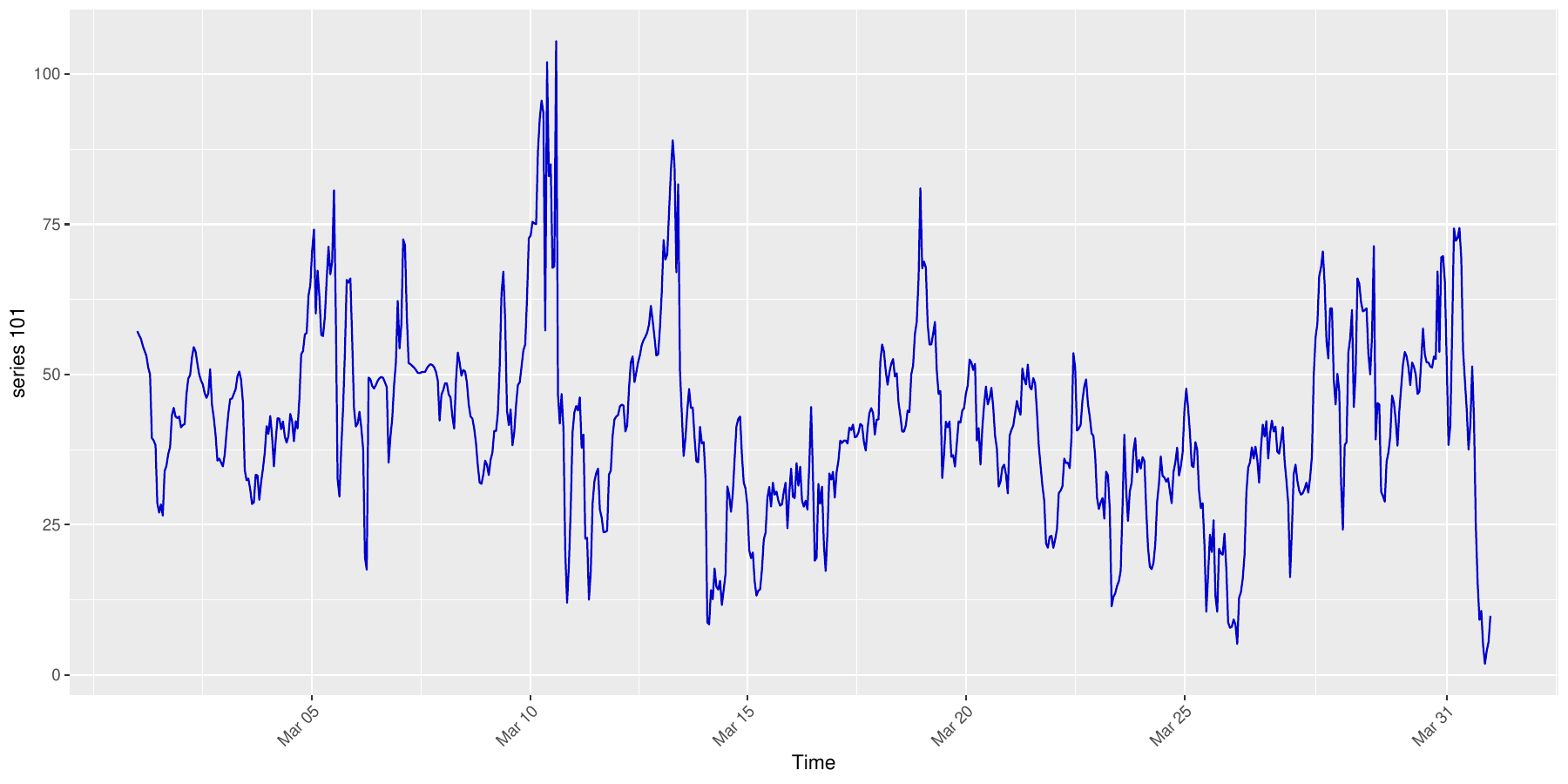}
        \caption{Series 101}
    \end{subfigure}
    \begin{subfigure}[t]{0.49\textwidth}
        \includegraphics[width=\linewidth]{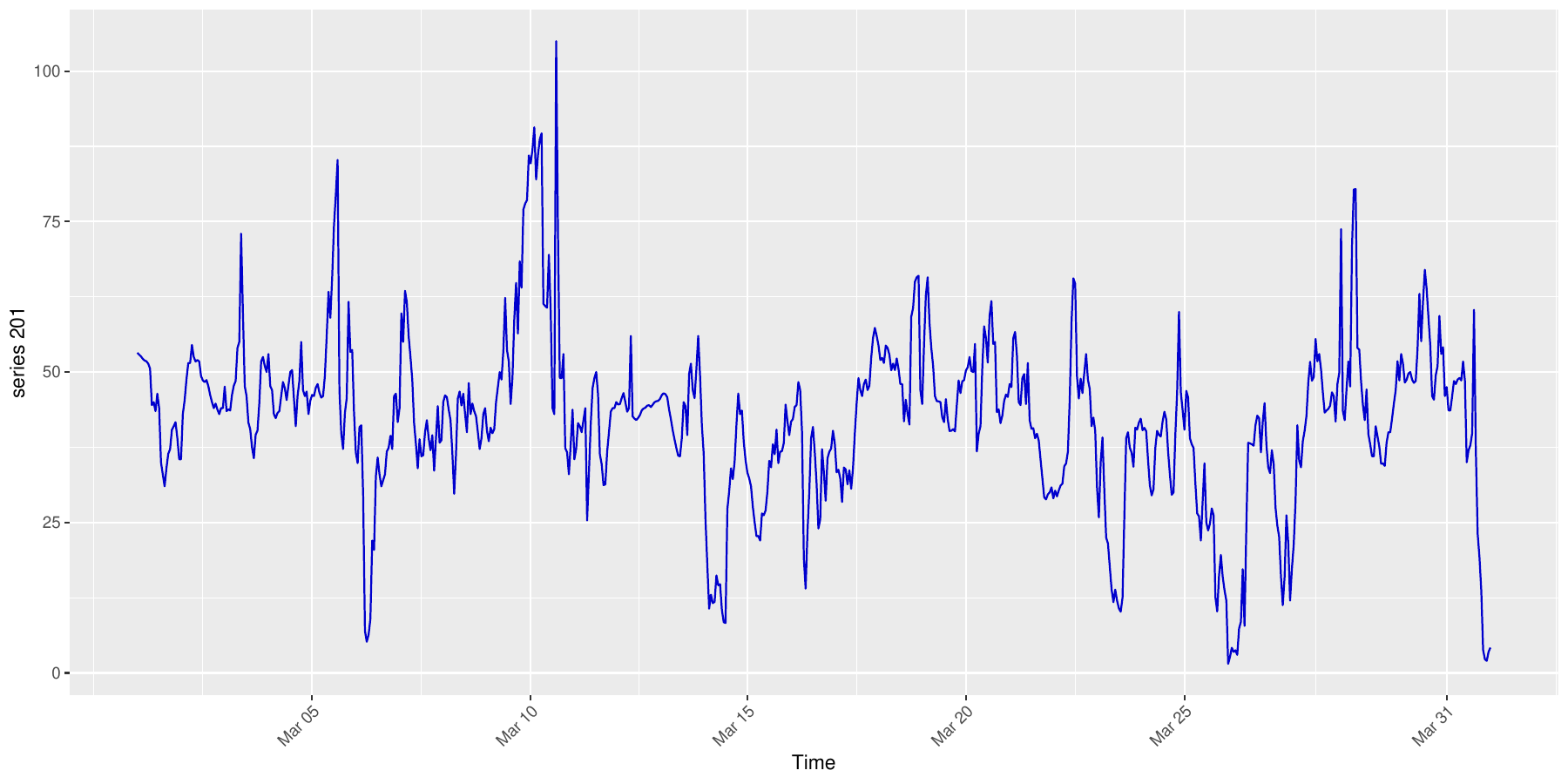}
        \caption{Series 201}
    \end{subfigure} \\
    \begin{subfigure}[t]{0.49\textwidth}
        \includegraphics[width=\linewidth]{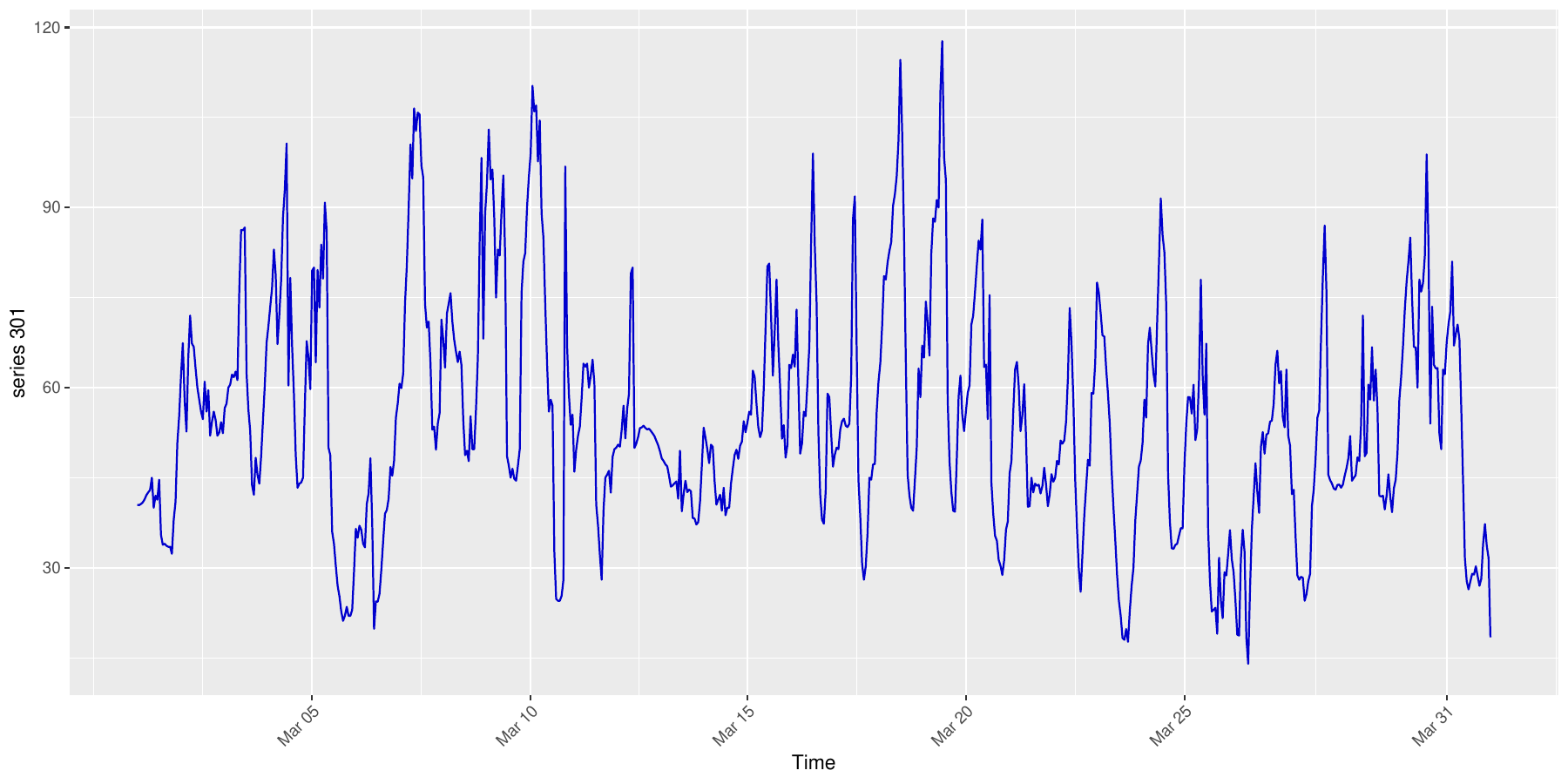}
        \caption{Series 301}
    \end{subfigure}
    \begin{subfigure}[t]{0.49\textwidth}
        \includegraphics[width=\linewidth]{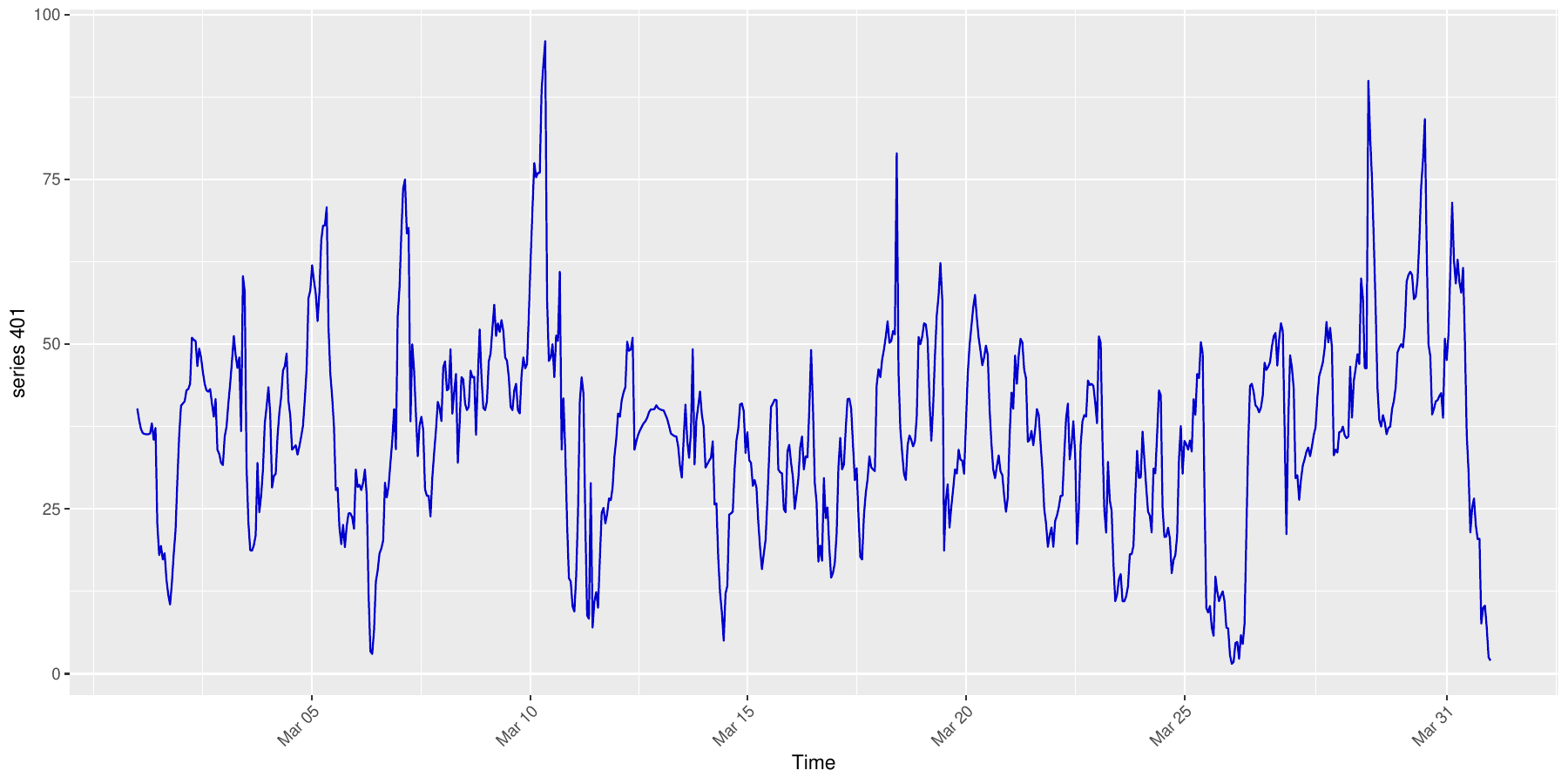}
        \caption{Series 401}
    \end{subfigure}
    \caption{Time plots of the four targeted hourly PM$_{2.5}$ measurements on Taiwan in March 2017.}
    \label{fig:2}
\end{figure}

We next consider the data of hourly PM$_{2.5}$ measurements in Taiwan during March of 2017.
The data are sourced from the \texttt{SLBDD} package \citep{slbdd} in \texttt{R}. 
Each series in the data set represents measurements (in micrograms per cubic meters, $\mathrm{\mu g/m^{3}}$) taken by a novel device known as the AirBox. 
After an initial examination of these series, we remove series 29 and 70 because these series are near identically zero except at a few time points, offering no useful variations.
Among the 516 series in the data set, we choose series 101, 201, 301, 401, as target series.
Figure \ref{fig:2} depicts their time plots. 

We consider both one-step-ahead ($h = 1$) and two-step-ahead ($h=2$) forecasts, and employ $q \in \{2,3\}$ lags and $r \in \{2, 4, 6\}$ factors in forecasting.
The last ten days of data (240 time periods) are reserved for out-of-sample testing.
Tables \ref{tab:series101}--\ref{tab:series401} present the results for $h= 1$. 
The proposed GO-sdPCA ranks as the most predictive method in terms of RMSFE for two of the four targeted series: Series 101 and 201. It outperforms the Lasso in forecasting all four targeted series.
The performance of LYB is also noteworthy. It ranks among the best two methods for all four targeted series.
The DM test, on the other hand, indicates the difference in forecasting accuracy is not statistically significant. 
Contrary to the case in the macroeconomic data, the proposed GsP$^{*}$ only outperformed the factor-based methods, including the sdPCA, with some weak confidence.
For the two-step-ahead forecasts, for which the results are reported in Tables \ref{tab:series101_2}--\ref{tab:series401_2}, the performance of various methods is more entangled, with the RF consistently ranked among the top two methods for three of the four targeted series.
With some weaker confidence, GsP$^{*}$ is the best performing factor-based method for two of the four targeted series. Again, the DM test fails to separate significantly various forecasting 
methods. We believe this result might be caused by the substantial uncertainty in the PM$_{2.5}$ 
measurements. 

In sum, for $h=1$, GsP$^{*}$ is effective in forecasting PM$_{2.5}$ data as well as the U.S. macroeconomic data.
This implies the proposed procedure is able to capture highly predictive factors across 
various applications.
On the other hand, for $h=2$, the dynamic nature of the data may be much more involved.
The performance of the proposed method becomes similar to those of the other forecasting methods. 

\begin{table}[h]
\centering
\caption{Root mean squared forecast errors for predicting series 101 in Taiwan PM$_{2.5}$ data. The forecast horizon $h$ is $1$. The lowest RMSFE achieved by each method is in boldface. Among these boldfaced values, the lowest two values are marked in red. $q$ denotes the number of lags used in estimation and $r$ is the number of factors extracted. GsP$^*$ denotes the proposed GO-sdPCA. The figures in the parentheses are the $p$-values of the Diebold-Mariano test of whether GsP$^{*}$ is more accurate.}
\begin{tabular}{@{}ccccccc@{}}
\toprule
      \multicolumn{7}{c}{Series 101 (RMSE); $h = 1$}       \\ 
$q$     & \multicolumn{3}{c}{2} & \multicolumn{3}{c}{3} \\ \cmidrule(lr){2-4} \cmidrule(lr){5-7} 
Lasso &  \multicolumn{3}{c}{6.363} & \multicolumn{3}{c}{\bf 6.327} \\
 &  \multicolumn{3}{c}{} & \multicolumn{3}{c}{(0.015)} \\
RF    &  \multicolumn{3}{c}{\bf 6.871} & \multicolumn{3}{c}{6.910} \\
 &  \multicolumn{3}{c}{(0.004)} & \multicolumn{3}{c}{} \\
$r$     & 2     & 4     & 6     & 2     & 4     & 6     \\ \cmidrule(lr){2-4} \cmidrule(lr){5-7} 
GsP$^*$ & 6.047  & 6.193  & 6.173 & \color{red}{\bf 6.021}  & 6.162  & 6.091 \\
sdPCA &  {\bf 6.079}  & 6.151  & 6.221   & 6.079  & 6.154  & 6.167  \\
& (0.298) & & & & & \\
SW    &  {\bf 6.098}  & 6.160  & 6.104   & 6.103  & 6.170  & 6.110  \\
& (0.269) & & & & & \\
LYB & \color{red}{\bf 6.055} & 6.125 & 6.117 & 6.060 & 6.133 & 6.146 \\
& (0.392) & & & & & \\
\bottomrule
\end{tabular}
\label{tab:series101}
\end{table}

\begin{table}[h]
\centering
\caption{Root mean squared forecast errors for predicting series 201 in Taiwan PM$_{2.5}$ data. The forecast horizon $h$ is $1$. The lowest RMSFE achieved by each method is in boldface. Among these boldfaced values, the lowest two values are marked in red. $q$ denotes the number of lags used in estimation and $r$ is the number of factors extracted. GsP$^*$ denotes the proposed GO-sdPCA. The figures in the parentheses are the $p$-values of the Diebold-Mariano test of whether GsP$^{*}$ is more accurate.}
\begin{tabular}{@{}ccccccc@{}}
\toprule
      \multicolumn{7}{c}{Series 201 (RMSE); $h = 1$}       \\ 
$q$     & \multicolumn{3}{c}{2} & \multicolumn{3}{c}{3} \\ \cmidrule(lr){2-4} \cmidrule(lr){5-7} 
Lasso &  \multicolumn{3}{c}{5.990} & \multicolumn{3}{c}{\bf 5.973} \\
&  \multicolumn{3}{c}{} & \multicolumn{3}{c}{(0.346)} \\
RF    &  \multicolumn{3}{c}{{\bf 5.938}} & \multicolumn{3}{c}{6.109} \\
&  \multicolumn{3}{c}{(0.399)} & \multicolumn{3}{c}{} \\
$r$     & \multicolumn{1}{c}{2} & \multicolumn{1}{c}{4} & \multicolumn{1}{c}{6} & \multicolumn{1}{c}{2} & \multicolumn{1}{c}{4} & \multicolumn{1}{c}{6} \\ \cmidrule(lr){2-4} \cmidrule(lr){5-7} 
GsP$^{*}$ & \color{red}{\bf 5.897}  & 6.079  & 6.128 & 5.931  & 6.197  & 6.146 \\
sdPCA &  6.012  & {\bf 5.982}  & 6.065   & 6.057  & 6.019  & 6.093  \\
& & (0.201) & & & & \\
SW    &  6.113  & 5.956  & {\bf 5.952}   & 6.135  & 5.989  & 5.984  \\
& & & (0.298) & & & \\
LYB & 6.129 & 5.967 & \color{red}{\bf 5.934} & 6.154 & 5.994 & 5.972 \\
& & & (0.358) & & & \\
\bottomrule
\end{tabular}
\label{tab:series201}
\end{table}

\begin{table}[h]
\centering
\caption{Root mean squared forecast errors for predicting series 301 in Taiwan PM$_{2.5}$ data. The forecast horizon $h$ is $1$. The lowest RMSFE achieved by each method is in boldface. Among these boldfaced values, the lowest two values are marked in red. $q$ denotes the number of lags used in estimation and $r$ is the number of factors extracted. GsP$^*$ denotes the proposed GO-sdPCA. The figures in the parentheses are the $p$-values of the Diebold-Mariano test of whether GsP$^{*}$ is more accurate.}
\begin{tabular}{@{}ccccccc@{}}
\toprule
      \multicolumn{7}{c}{Series 301 (RMSE); $h = 1$}       \\ 
$q$     & \multicolumn{3}{c}{2} & \multicolumn{3}{c}{3} \\ \cmidrule(lr){2-4} \cmidrule(lr){5-7} 
Lasso &  \multicolumn{3}{c}{\bf 6.824} & \multicolumn{3}{c}{6.854} \\
 &  \multicolumn{3}{c}{(0.367)} & \multicolumn{3}{c}{} \\
RF    &  \multicolumn{3}{c}{\color{red}{\bf 6.653}} & \multicolumn{3}{c}{6.698} \\
 &  \multicolumn{3}{c}{(0.601)} & \multicolumn{3}{c}{} \\
$r$     & \multicolumn{1}{c}{2} & \multicolumn{1}{c}{4} & \multicolumn{1}{c}{6} & \multicolumn{1}{c}{2} & \multicolumn{1}{c}{4} & \multicolumn{1}{c}{6} \\ \cmidrule(lr){2-4} \cmidrule(lr){5-7} 
GsP$^*$ & 6.756  & 6.765  & {\bf 6.724} & 6.800  & 6.813  & 6.781 \\
sdPCA & {\bf 6.733}  & 6.977  & 6.865   & 6.757  & 7.021  & 6.879  \\
& (0.483) & & & & & \\
SW    &  6.794  & 7.029  & 7.040   & {\bf 6.743}  & 7.013  & 7.009  \\
& & & & (0.476) & & \\
LYB   & 6.749 & 6.966 & 7.018 & \color{red}{{\bf 6.710}} & 6.958 & 7.047\\
& & & & (0.516) & & \\
\bottomrule
\end{tabular}
\label{tab:series301}
\end{table}

\begin{table}[h]
\centering
\caption{Root mean squared forecast errors for predicting series 401 in Taiwan PM$_{2.5}$ data. The forecast horizon $h$ is $1$. The lowest RMSFE achieved by each method is in boldface. Among these boldfaced values, the lowest two values are marked in red. $q$ denotes the number of lags used in estimation and $r$ is the number of factors extracted. GsP$^*$ denotes the proposed GO-sdPCA. The figures in the parentheses are the $p$-values of the Diebold-Mariano test of whether GsP$^{*}$ is more accurate.}
\begin{tabular}{@{}ccccccc@{}}
\toprule
      \multicolumn{7}{c}{Series 401 (RMSE); $h = 1$}       \\ 
$q$     & \multicolumn{3}{c}{2} & \multicolumn{3}{c}{3} \\ \cmidrule(lr){2-4} \cmidrule(lr){5-7} 
Lasso &  \multicolumn{3}{c}{6.392} & \multicolumn{3}{c}{\bf 6.349} \\
&  \multicolumn{3}{c}{} & \multicolumn{3}{c}{(0.125)} \\
RF    &  \multicolumn{3}{c}{\bf 7.122} & \multicolumn{3}{c}{7.196} \\
&  \multicolumn{3}{c}{(0.008)} & \multicolumn{3}{c}{} \\
$r$     & \multicolumn{1}{c}{2} & \multicolumn{1}{c}{4} & \multicolumn{1}{c}{6} & \multicolumn{1}{c}{2} & \multicolumn{1}{c}{4} & \multicolumn{1}{c}{6} \\ \cmidrule(lr){2-4} \cmidrule(lr){5-7} 
GsP$^*$ & {\bf 6.170}  & 6.235  & 6.251 & 6.238  & 6.236  & 6.243 \\
sdPCA & 6.200   & 6.026  & 6.111   & 6.272  & {\bf 6.016}  & 6.112  \\
& & & & & (0.870) & \\
SW    & 6.171   & \color{red}{\bf 5.908}  & 5.964   & 6.227  & 5.953  & 5.999  \\
& & (0.996) & & & & \\
LYB   & 6.139   & \color{red}{\bf 5.928} & 5.994 & 6.193 & 5.967 & 6.023 \\
& & (0.992) & & & & \\
\bottomrule
\end{tabular}
\label{tab:series401}
\end{table}

\begin{table}[h]
\centering
\caption{Root mean squared forecast errors for predicting series 101 in Taiwan PM$_{2.5}$ data. The forecast horizon $h$ is $2$. The lowest RMSFE achieved by each method is in boldface. Among these boldfaced values, the lowest two values are marked in red. $q$ denotes the number of lags used in estimation and $r$ is the number of factors extracted. GsP$^*$ denotes the proposed GO-sdPCA. The figures in the parentheses are the $p$-values of the Diebold-Mariano test of whether GsP$^{*}$ is more accurate.}
\begin{tabular}{@{}crrrrrr@{}}
\toprule
      \multicolumn{7}{c}{Series 101 (RMSE); $h = 2$}       \\ 
$q$     & \multicolumn{3}{c}{2} & \multicolumn{3}{c}{3} \\ \cmidrule(lr){2-4} \cmidrule(lr){5-7} 
Lasso &  \multicolumn{3}{c}{\color{red}{\bf 9.584}} & \multicolumn{3}{c}{9.618} \\
&  \multicolumn{3}{c}{(0.818)} & \multicolumn{3}{c}{} \\
RF    &  \multicolumn{3}{c}{9.544} & \multicolumn{3}{c}{\color{red}{\bf 9.415}} \\
&  \multicolumn{3}{c}{} & \multicolumn{3}{c}{(0.803)} \\
$r$     & \multicolumn{1}{c}{2} & \multicolumn{1}{c}{4} & \multicolumn{1}{c}{6} & \multicolumn{1}{c}{2} & \multicolumn{1}{c}{4} & \multicolumn{1}{c}{6} \\ \cmidrule(lr){2-4} \cmidrule(lr){5-7} 
GsP$^*$ & 9.788  & 9.969  & 9.915 & {\bf 9.758}  & 9.941  & 9.893 \\
sdPCA &  9.982  & 9.970  & 9.804   & 9.980  & 9.930  & {\bf 9.715}  \\
& & & & & & (0.597) \\
SW    &  10.073  & 9.946  & {\bf 9.868}   & 10.069  & 9.951  & 9.875  \\
& & & (0.326) & & & \\
LYB   &  10.000  & 9.888  & {\bf 9.883} & 9.994 & 9.888 & 9.907 \\
& & & (0.282) & & & \\
\bottomrule
\end{tabular}
\label{tab:series101_2}
\end{table}

\begin{table}[h]
\centering
\caption{Root mean squared forecast errors for predicting series 201 in Taiwan PM$_{2.5}$ data. The forecast horizon $h$ is $2$. The lowest RMSFE achieved by each method is in boldface. Among these boldfaced values, the lowest two values are marked in red. $q$ denotes the number of lags used in estimation and $r$ is the number of factors extracted. GsP$^*$ denotes the proposed GO-sdPCA. The figures in the parentheses are the $p$-values of the Diebold-Mariano test of whether GsP$^{*}$ is more accurate.}
\begin{tabular}{@{}crrrrrr@{}}
\toprule
      \multicolumn{7}{c}{Series 201 (RMSE); $h = 2$}       \\ 
$q$     & \multicolumn{3}{c}{2} & \multicolumn{3}{c}{3} \\ \cmidrule(lr){2-4} \cmidrule(lr){5-7} 
Lasso &  \multicolumn{3}{c}{\color{red}{\bf 9.868}} & \multicolumn{3}{c}{10.005} \\
&  \multicolumn{3}{c}{(0.731)} & \multicolumn{3}{c}{} \\
RF    &  \multicolumn{3}{c}{\color{red}{\bf 9.594}} & \multicolumn{3}{c}{9.614} \\
&  \multicolumn{3}{c}{(0.987)} & \multicolumn{3}{c}{} \\
$r$     & \multicolumn{1}{c}{2} & \multicolumn{1}{c}{4} & \multicolumn{1}{c}{6} & \multicolumn{1}{c}{2} & \multicolumn{1}{c}{4} & \multicolumn{1}{c}{6} \\ \cmidrule(lr){2-4} \cmidrule(lr){5-7} 
GsP$^*$ & {\bf 9.998}  & 10.276  & 10.209 & 10.052  & 10.292  & 10.219 \\
sdPCA &  10.164  & {\bf 10.070}  & 10.112  & 10.228  & 10.154  & 10.150  \\
& & (0.339) & & & & \\
SW    &  10.233  & {\bf 10.151}  & 10.151   & 10.270  & 10.196  & 10.207  \\
& & (0.244) & & & & \\
LYB   &  10.230  & 10.186  & {\bf 10.176}  & 10.267  & 10.222  &  10.269  \\
& & & (0.212) & & & \\
\bottomrule
\end{tabular}
\label{tab:series201_2}
\end{table}

\begin{table}[h]
\centering
\caption{Root mean squared forecast errors for predicting series 301 in Taiwan PM$_{2.5}$ data. The forecast horizon $h$ is $2$. The lowest RMSFE achieved by each method is in boldface. Among these boldfaced values, the lowest two values are marked in red. $q$ denotes the number of lags used in estimation and $r$ is the number of factors extracted. GsP$^*$ denotes the proposed GO-sdPCA. The figures in the parentheses are the $p$-values of the Diebold-Mariano test of whether GsP$^{*}$ is more accurate.}
\begin{tabular}{@{}crrrrrr@{}}
\toprule
      \multicolumn{7}{c}{Series 301 (RMSE); $h = 2$}       \\ 
$q$     & \multicolumn{3}{c}{2} & \multicolumn{3}{c}{3} \\ \cmidrule(lr){2-4} \cmidrule(lr){5-7} 
Lasso &  \multicolumn{3}{c}{11.189} & \multicolumn{3}{c}{\bf 10.968} \\
&  \multicolumn{3}{c}{} & \multicolumn{3}{c}{(0.314)} \\
RF    &  \multicolumn{3}{c}{\color{red}{\bf 10.727}} & \multicolumn{3}{c}{10.734} \\
&  \multicolumn{3}{c}{(0.542)} & \multicolumn{3}{c}{} \\
$r$     & \multicolumn{1}{c}{2} & \multicolumn{1}{c}{4} & \multicolumn{1}{c}{6} & \multicolumn{1}{c}{2} & \multicolumn{1}{c}{4} & \multicolumn{1}{c}{6} \\ \cmidrule(lr){2-4} \cmidrule(lr){5-7} 
GsP$^*$ & 10.895  & 10.899  & 10.890  & 10.828  & 10.838  & \color{red}{\bf 10.767} \\
sdPCA & 10.930  & 11.308  & 11.220   & {\bf 10.794}  & 11.164  & 11.140  \\
& & & & (0.460) & & \\
SW    & 11.265   & 11.675  & 11.598   & {\bf 11.038}  & 11.528  & 11.456  \\
& & & & (0.280) & & \\
LYB   & 11.130   & 11.448  & 11.442   & {\bf 10.930}   & 11.325  & 11.325 \\
& & & & (0.362) & & \\
\bottomrule
\end{tabular}
\label{tab:series301_2}
\end{table}

\begin{table}[h]
\centering
\caption{Root mean squared forecast errors for predicting series 401 in Taiwan PM$_{2.5}$ data. The forecast horizon $h$ is $2$. The lowest RMSFE achieved by each method is in boldface. Among these boldfaced values, the lowest two values are marked in red. $q$ denotes the number of lags used in estimation and $r$ is the number of factors extracted. GsP$^*$ denotes the proposed GO-sdPCA. The figures in the parentheses are the $p$-values of the Diebold-Mariano test of whether GsP$^{*}$ is more accurate.}
\begin{tabular}{@{}ccccccc@{}}
\toprule
      \multicolumn{7}{c}{Series 401 (RMSE); $h = 2$}       \\ 
$q$     & \multicolumn{3}{c}{2} & \multicolumn{3}{c}{3} \\ \cmidrule(lr){2-4} \cmidrule(lr){5-7} 
Lasso &  \multicolumn{3}{c}{\bf 9.862} & \multicolumn{3}{c}{9.965} \\
&  \multicolumn{3}{c}{(0.100)} & \multicolumn{3}{c}{} \\
RF    &  \multicolumn{3}{c}{9.694} & \multicolumn{3}{c}{\bf 9.676} \\
&  \multicolumn{3}{c}{} & \multicolumn{3}{c}{(0.323)} \\
$r$     & \multicolumn{1}{c}{2} & \multicolumn{1}{c}{4} & \multicolumn{1}{c}{6} & \multicolumn{1}{c}{2} & \multicolumn{1}{c}{4} & \multicolumn{1}{c}{6} \\ \cmidrule(lr){2-4} \cmidrule(lr){5-7} 
GsP$^*$ & 9.768    & 9.572  & 9.668    & 9.745   & {\bf 9.491}  & 9.626 \\
sdPCA   & 10.391   & 9.317  & 9.427    & 10.388  & \color{red}{\bf 9.213}  & 9.371  \\
& & & & & (0.891) & \\
SW      & 10.405   & 9.570  & 9.200    & 10.401  & 9.508  & \color{red}{\bf 9.172}  \\
& & & & & & (0.798) \\
LYB     & 10.293   & 9.600  & 9.285    & 10.278  & 9.534  & {\bf 9.279} \\
& & & & & & (0.747) \\
\bottomrule
\end{tabular}
\label{tab:series401_2}
\end{table}

\section{Discussion and concluding remarks}
In this work, we proposed a novel method for time series forecasting when many predictors are available.
The rationale behind our method is to mine the possibly many (compared to the sample size) factor-relevant predictors while reducing the effect of the many irrelevant variables in the high-dimensional data.
The results in the simulation studies and the empirical applications suggest that the proposed method is useful for improving upon both the factor-based methods and methods for sparse linear models such as the Lasso.
Finally, we remark that the theoretical investigation of the peeling technique, a key ingredient in our method, would be an interesting topic for future research.

\clearpage
\bibliographystyle{apalike}
\bibliography{thebib.bib}

\begin{thebibliography}{}

\bibitem[Barron et~al., 2008]{barron2008}
Barron, A.~R., Cohen, A., Dahmen, W., and DeVore, R.~A. (2008).
\newblock Approximation and learning by greedy algorithms.
\newblock {\em The Annals of Statistics}, 36(1):64--94.

\bibitem[Bernanke and Boivin, 2003]{Bernanke2003}
Bernanke, B.~S. and Boivin, J. (2003).
\newblock Monetary policy in a data-rich environment.
\newblock {\em Journal of Monetary Economics}, 50(3):525--546.

\bibitem[Bernanke et~al., 2005]{Bernanke2005}
Bernanke, B.~S., Boivin, J., and Eliasz, P. (2005).
\newblock {Measuring the Effects of Monetary Policy: A Factor-Augmented Vector Autoregressive (FAVAR) Approach}.
\newblock {\em The Quarterly Journal of Economics}, 120(1):387--422.

\bibitem[Boivin and Ng, 2006]{Boivin2006}
Boivin, J. and Ng, S. (2006).
\newblock Are more data always better for factor analysis?
\newblock {\em Journal of Econometrics}, 132(1):169--194.

\bibitem[Breiman, 2001]{Breiman2001}
Breiman, L. (2001).
\newblock Random forests.
\newblock {\em Machine Learning}, 45(1):5--32.

\bibitem[Caro et~al., 2022]{slbdd}
Caro, A., Elias, A., Peña, D., and Tsay, R.~S. (2022).
\newblock {\em SLBDD: Statistical Learning for Big Dependent Data}.
\newblock R package version 0.0.4.

\bibitem[Chan et~al., 2017]{chan2017}
Chan, N.~H., Ing, C.-K., Li, Y., and Yau, C.~Y. (2017).
\newblock Threshold estimation via group orthogonal greedy algorithm.
\newblock {\em Journal of Business \& Economic Statistics}, 35(2):334--345.

\bibitem[Chi et~al., 2022]{Chi2022}
Chi, C.-M., Vossler, P., Fan, Y., and Lv, J. (2022).
\newblock {Asymptotic properties of high-dimensional random forests}.
\newblock {\em The Annals of Statistics}, 50(6):3415--3438.

\bibitem[Diebold and Mariano, 1995]{Diebold1995}
Diebold, F.~X. and Mariano, R.~S. (1995).
\newblock Comparing predictive accuracy.
\newblock {\em Journal of Business \& Economic Statistics}, 13(3):253--263.

\bibitem[Gao and Tsay, 2024]{gao2023}
Gao, Z. and Tsay, R.~S. (2024+).
\newblock Supervised dynamic pca: Linear dynamic forecasting with many predictors.
\newblock {\em Journal of the American Statistical Association, (accepted).}

\bibitem[Harvey et~al., 1997]{HARVEY1997}
Harvey, D., Leybourne, S., and Newbold, P. (1997).
\newblock Testing the equality of prediction mean squared errors.
\newblock {\em International Journal of Forecasting}, 13(2):281--291.

\bibitem[Huang et~al., 2022]{Huang2022}
Huang, D., Jiang, F., Li, K., Tong, G., and Zhou, G. (2022).
\newblock Scaled pca: A new approach to dimension reduction.
\newblock {\em Management Science}, 68(3):1678--1695.

\bibitem[Ing, 2020]{ing2020}
Ing, C.-K. (2020).
\newblock Model selection for high-dimensional linear regression with dependent observations.
\newblock {\em The Annals of Statistics}, 48(4):1959--1980.

\bibitem[Ing and Lai, 2011]{ing2011}
Ing, C.-K. and Lai, T.~L. (2011).
\newblock A stepwise regression method and consistent model selection for high-dimensional sparse linear models.
\newblock {\em Statistica Sinica}, 21(4):1473--1513.

\bibitem[James et~al., 2021]{james2021}
James, G., Witten, D., Hastie, T., and Tibshirani, R. (2021).
\newblock {\em An Introduction to Statistical Learning with Appications in R}.
\newblock Springer New York, NY, second edition.

\bibitem[Jurado et~al., 2015]{Jurado2015}
Jurado, K., Ludvigson, S.~C., and Ng, S. (2015).
\newblock Measuring uncertainty.
\newblock {\em The American Economic Review}, 105(3):1177--1216.

\bibitem[Kim and Swanson, 2014]{Kim2014}
Kim, H.~H. and Swanson, N.~R. (2014).
\newblock Forecasting financial and macroeconomic variables using data reduction methods: New empirical evidence.
\newblock {\em Journal of Econometrics}, 178:352--367.

\bibitem[Lam et~al., 2011]{lam2011}
Lam, C., Yao, Q., and Bathia, N. (2011).
\newblock Estimation of latent factors for high-dimensional time series.
\newblock {\em Biometrika}, 98(4):901--918.

\bibitem[Liaw and Wiener, 2002]{randomForest}
Liaw, A. and Wiener, M. (2002).
\newblock Classification and regression by randomforest.
\newblock {\em R News}, 2(3):18--22.

\bibitem[McCracken and Ng, 2016]{maccraken2016}
McCracken, M.~W. and Ng, S. (2016).
\newblock Fred-md: A monthly database for macroeconomic research.
\newblock {\em Journal of Business \& Economic Statistics}, 34(4):574--589.

\bibitem[Medeiros and Mendes, 2016]{medeiros2016}
Medeiros, M.~C. and Mendes, E.~F. (2016).
\newblock $\ell_{1}$-regularization of high-dimensional time-series models with non-gaussian and heteroskedastic errors.
\newblock {\em Journal of Econometrics}, 191(1):255--271.

\bibitem[Pe\~{n}a and Box, 1987]{pena1987}
Pe\~{n}a, D. and Box, G.~E. (1987).
\newblock Identifying a simplifying structure in time series.
\newblock {\em Journal of the American statistical Association}, 82(399):836--843.

\bibitem[Saha et~al., 2023]{Saha2023}
Saha, A., Basu, S., and Datta, A. (2023).
\newblock Random forests for spatially dependent data.
\newblock {\em Journal of the American Statistical Association}, 118(541):665--683.

\bibitem[Stock and Watson, 2002a]{SW2002}
Stock, J.~H. and Watson, M.~W. (2002a).
\newblock Forecasting using principal components from a large number of predictors.
\newblock {\em Journal of the American Statistical Association}, 97(460):1167--1179.

\bibitem[Stock and Watson, 2002b]{SW2002b}
Stock, J.~H. and Watson, M.~W. (2002b).
\newblock Macroeconomic forecasting using diffusion indexes.
\newblock {\em Journal of Business \& Economic Statistics}, 20(2):147--162.

\bibitem[Temlyakov, 2011]{Temlyakov2011}
Temlyakov, V. (2011).
\newblock {\em Greedy Approximation}.
\newblock Cambridge Monographs on Applied and Computational Mathematics. Cambridge University Press.

\bibitem[Tibshirani, 1996]{tib1996}
Tibshirani, R. (1996).
\newblock Regression shrinkage and selection via the lasso.
\newblock {\em Journal of the Royal Statistical Society. Series B (Methodological)}, 58(1):267--288.

\bibitem[Tsay, 2013]{tsay2013multivariate}
Tsay, R.~S. (2013).
\newblock {\em Multivariate time series analysis: with R and financial applications}.
\newblock John Wiley \& Sons.

\end{thebibliography}

\end{document}